\documentclass{aa} 

\usepackage{hyperref}
\usepackage{graphicx}
\usepackage{txfonts}
\usepackage{orcidlink}
\usepackage{placeins}
\usepackage{xspace}  

\newcommand{\HI}{H\,\textsc{i}\xspace}
\newcommand{\Msol}{$\rm{M_{\odot}}$\xspace}
\newcommand{\Mstar}{$M_\mathrm{\star}$\xspace}
\newcommand{\MHI}{$M_\mathrm{HI}$\xspace}
\newcommand{\NHI}{$N_\mathrm{HI}$\xspace}
\newcommand{\SigHII}{$\Sigma_\mathrm{H2}$\xspace}
\newcommand{\rT}{$r_\mathrm{t}$\xspace}
\newcommand{\kms}{km\,s$^{-1}$\xspace}
\newcommand{\cm}{cm$^{-2}$\xspace}

\newcommand{\Mdyn}{$M_\mathrm{dyn}$\xspace}

\newcommand{\vsys}{$v_{\rm sys}$\xspace}
\newcommand{\FHI}{f$_{\rm HI}$\xspace}

\newcommand{\Ha}{H$\alpha$\xspace}
\newcommand{\degree}{$^\circ$\xspace}
\newcommand{\SoFiA}{\texttt{SoFiA-2}\xspace}
\newcommand{\BBarolo}{\texttt{$^{\mathrm{3D}}$BAROLO}\xspace}

\begin{document} 

   \title{The origin and distribution of diffuse \HI in the close interacting MHONGOOSE galaxy systems}

   \titlerunning{Diffuse \HI in close systems}

   \author{D.~Kleiner\inst{1, 2}\orcidlink{0000-0002-7573-555X}, W.~J.~G.~de Blok\inst{1, 3 ,4}\orcidlink{0000-0001-8957-4518}, J.~Healy\inst{5, 6}\orcidlink{0000-0003-1020-8684}, F.~M.~Maccagni\inst{2, 7}\orcidlink{0000-0002-9930-1844}, E.~M.~Di~Teodoro\inst{8, 9}\orcidlink{0000-0003-4019-0673}, A.~Fraser-McKelvie\inst{10}\orcidlink{0000-0001-9557-5648}, A.~Marasco\inst{11}, S.~Veronese\inst{12}\orcidlink{0000-0003-4789-9685}, \and S.~Kurapati\inst{13}\orcidlink{0000-0001-6615-5492}}
   \authorrunning{Kleiner et al.}

  \institute{Netherlands Institute for Radio Astronomy (ASTRON), Oude Hoogeveensedijk 4, 7991 PD Dwingeloo, the Netherlands\\
              \email{kleiner@astron.nl}
              \and
              INAF -- Osservatorio Astronomico di Cagliari, Via della Scienza 5, I-09047 Selargius (CA), Italy
              \and
              Department of Astronomy, University of Cape Town, Private Bag X3, Rondebosch 7701, South Africa
              \and
              Kapteyn Astronomical Institute, University of Groningen, PO Box 800, NL-9700 AV Groningen, the Netherlands
              \and
              Jodrell Bank Centre for Astrophysics, School of Physics and Astronomy, University of Manchester, Oxford Road, Manchester M13 9PL, UK
              \and
              Department of Astronomy, University of Cape Town, Private Bag X3, Rondebosch 7701, South Africa
              \and
              Wits Centre for Astrophysics, School of Physics, University of the Witwatersrand, 1 Jan Smuts Avenue, 2000, Johannesburg, South Africa
              \and
              Universit\`{a} di Firenze, Dipartimento di Fisica e Astronomia, Largo Enrico Fermi 2, 50125, Firenze, Italy
              \and
              INAF - Osservatorio Astrofisico di Arcetri, Largo E. Fermi 5, 50125 Firenze, Italy
              \and 
             European Southern Observatory, Karl-Schwarzschild-Straße 2, Garching, 85748, Germany
             \and 
             INAF -- Padova Astronomical Observatory, Vicolo dell’Osservatorio 5, I-35122 Padova, Italy
             \and 
             Max Planck Institute for Radio Astronomy, Auf dem Hügel 69, 53121 Bonn, Germany
             \and
             National Centre for Radio Astrophysics, Tata Institute of Fundamental Research (NCRA-TIFR), Pune 411007, India}
   \date{Received 10 June 2026 / Accepted 23 October 2026}

  \abstract{We present the deepest \HI observations of two systems with $D_\mathrm{proj} \leq 100$\,kpc and $\Delta v_\mathrm{sys} \leq 200$\,\kms in the MHONGOOSE survey. The MeerKAT observations are characterised by 3$\sigma$ \NHI = 10$^{17.9}$\,\cm over 16\,\kms at 4\,kpc scales. The new data have enabled us to detect \HI in seven galaxies: the interacting NGC\,3511/3513 pair in one field and an interacting triplet (NGC\,4781, NGC\,4790 and UGCA\,308), along with two new (2MASX\,J1255 and MKT\,25413) dwarf satellites in the second field. Six galaxies lie on the star-forming main sequence and their \HI gas fractions are typical for such systems. In contrast, MKT\,25413 exhibits log(\MHI/\Msol) = 5.7, which is the lowest \MHI detected of a galaxy that is not a Milky Way satellite. This low-surface-brightness dwarf galaxy shows an anomalously low level of star formation and low \HI fraction for its stellar mass. We also found a ubiquitous low column density \HI surrounding the galaxies, which is a likely consequence of mild tidal interactions from fly-bys in the past few hundred megayears. We derive the stellar and \Ha kinematics from VLT/MUSE observations and the H$_{2}$ kinematics from ALMA observations. The kinematics of the stellar and multiphase (\HI, H$_{2}$ and \Ha) interstellar medium of the interacting galaxies reveal an ordered rotation and the atomic-to-molecular gas ratios are indistinguishable from star-forming field galaxies. The stellar morphology of all the interacting galaxies (apart from the low-mass dwarf UGCA\,308) does not show any major disruptions and there is no stellar component associated with the \HI tails, bridges or clouds in the intra-group mediums, down to $\mu_{g}$ = 29\,mag\,arcsec$^{-2}$. We find more \HI beyond the tidal radius of the galaxies with the more recent fly-bys and we surmise that the mild environmental interactions in the close systems have the capacity to pull the diffuse \HI out of the disc and redistribute it. The gas is not in equilibrium, although it remains bound and is in the process of falling back onto the disc.}


   \maketitle

\section{Introduction}
The availability of a gas reservoir underpins the evolution of galaxies. When there is a sufficient gas supply, under the right conditions the gas can cool and further collapse to form stars \citep[e.g][]{Kennicutt1994, Kennicutt1998, Schinnerer2024}. If the gas reservoir has been removed or the available gas is unable to cool, star formation (SF) ceases, and the galaxy will be quenched \citep[e.g. see][and references within]{Cortese2021}. By using more sensitive observations and refined simulations, we are able to formulate a detailed picture of the baryon cycle that traces the gas and SF processes in and around galaxies.  

There has been extensive observational and theoretical work showing that gas from the interstellar medium (ISM) can be heated or ejected via different internal processes. For example, stellar winds, radiation, or supernova explosions can prevent cooling \citep[e.g.][]{Hopkins2012, Strickland2000, Chevance2022}. This feedback is most efficient in galaxies below log(\Mstar/\Msol) = 10, whereas galaxies above this mass are more likely to have active galactic nuclei (AGN) injecting energy into the ISM and thereby heating the surrounding gas and creating outflows \cite[e.g.][]{Croton2006, Fabian2012, Pillepich2018}. Another pathway to removing the gas reservoir of a galaxy is through external interactions that are directly related to the galactic ecosystem. Gas stripping efficiency is correlated with the galaxy environment (i.e. with the galaxy number density), while ram pressure appears to be the dominant and most efficient method to strip the ISM in dense galaxy cluster environments \citep[e.g.][]{Chung2007, Jaffe2018, Rohr2023, Edler2026}. The influence of tidal fields as galaxies orbit each other in close proximity has shown to profoundly disturb and sometimes strip the ISM in a wide range of environments \citep[e.g.][]{deBlok2018, Kleiner2021, Serra2023, Rhee2026}. Tidal fields can strip the stars and ISM in galaxy pairs, as well as low-density groups and clusters. It is also thought to be the main quenching mechanism in galaxy groups due to their cooler intra-group medium (IGM), lower orbital velocities, and the long-lasting encounters that result from these effects \citep[e.g.][]{Wu2022, Wang2022, Lin2023}. As tidal fields operate over a broad range of strengths, there are many factors that determine if the stellar component is affected and whether the ISM is disturbed or completely stripped. The situation is even less clear for galaxy pairs. In some pairs, tidal torques funnel the gas inward and boost central SF \citep{Barnes1996, Mihos1996}, resulting in enhanced atomic and molecular gas fractions \citep[e.g.][]{Violino2018, Ellison2018}, consistent with tidal tails channelling diffuse gas that condenses out of the surrounding hot halo \citep[e.g.][]{Wang2023}. Other studies have reported atomic gas stripping and SF suppression in pairs \citep[e.g.][]{Yu2022}. Both enhancement and suppression are seen even within the same interacting samples, depending on pair separation \citep[e.g.][]{Siljeg2025}. Galaxy pairs show larger scatter in gas content than isolated galaxies at fixed stellar mass, with comparable median values \citep[e.g.][]{Bok2020}, indicating that whether a given pair is gas-rich and star-forming or gas-poor and quenched likely depends on factors such as separation, mass ratio, and orbital stage.

In the low-$z$ Universe, the dominant component of the multiphase ISM is neutral atomic hydrogen (\HI), which has historically been shown to be one of the best observational tracers for ISM perturbations and stripping. As the interaction strength determines whether the gas reservoir is removed from the galaxy, resolved observations of \HI have provided many insights into the quenching stage of interacting galaxies \citep[e.g.][]{Koribalski2004a, Boselli2006, deBlok2018, Pearson2018, Luber2022}. Over the past two decades, a number of \HI interferometric surveys were carried out to observe the ISM and extended gaseous disc of nearby star-forming galaxies. Two important examples are the \HI Nearby Galaxy Survey \citep[THINGS;][]{Walter2008} with the Very Large Array (VLA) and Hydrogen Accretion in LOcal GAlaxieS \citep[HALOGAS;][]{Heald2011} with the Westerbork Synthesis Radio Telescope (WSRT). The 3$\sigma$ \HI column density (\NHI) -- normalised to a velocity width of 16\,\kms -- of THINGS is sensitive down to \NHI = 10$^{19.5}$\,\cm at a spatial resolution of a few hundred parsecs. It is focussed mainly on the kinematics and star-forming process of the dense \HI in the disc \citep[e.g][]{deBlok2008, Leroy2008, Bigiel2010}. On the other hand, HALOGAS is sensitive down to \NHI = 10$^{19}$\,\cm) over the same velocity width, at a spatial resolution of approximately one kiloparsec. It was able to detect and quantify the behaviour of extra-planar \HI in nearby gas-rich, star-forming galaxies \citep[e.g.][]{deBlok2014, Vargas2017, Marasco2019}. While these surveys were able to measure the \HI properties of galaxies in a column density regime approximately two orders of magnitude lower than where SF is known to occur, they were limited to the higher density \HI that only moderate-to-strong interactions are able to disrupt. Tidal interactions in groups that strip \HI above \NHI $\sim$ 10$^{19}$\,\cm are typically strong enough to produce massive (e.g. log(\MHI/\Msol) $>$ 8.5) \HI tails, and disturb the stellar component \citep[e.g.][]{Koribalski2005, Serra2013, Saponara2018}. 

Since MeerKAT \citep{Camilo2018} has come online, it has become possible to routinely observe \HI with sub-kpc resolution below \NHI = 10$^{19}$\,\cm\footnote{For an overview of \HI surveys and their respective resolution and sensitivities, see \citet{Maccagni2024b}.} at z $\sim$ 0. This combination has opened up a new parameter-space of resolved, low-column density \HI in nearby star-forming galaxies to explore. Not only can we determine the extent and mass contribution, we can quantify the effect of internal and environmental processes on the loosely bound, low-column-density \HI. This is very useful for tidal interactions in group environments, which is one of the least constrained mechanisms of the baryon cycle. The MeerKAT \HI Observations of Nearby Galactic Objects: Observing Southern Emitters \citep[MHONGOOSE;][]{deBlok2024} is a deep, MeerKAT \HI survey that has observed 30 nearby gas-rich, star-forming galaxies ranging between log(\MHI/\Msol) $\approx$ 7 -- 11, in isolated and low-density environments. The MHONGOOSE survey has produced the deepest, resolved \HI observations of nearby galaxies to date, reaching column densities below \NHI $\sim$ 10$^{18}$\,\cm over a velocity width of 16\,\kms. This probes scales of a few kiloparsecs, approximately 1.5 orders of magnitude lower than previous interferometric surveys (e.g. THINGS and HALOGAS) with a comparable spatial resolution. Exploiting this new parameter-space, MHONGOOSE observations have been able to detect analogues of local group satellites at tens of Mpc \citep{Maccagni2024a}, test theories of cold gas accretion \citep{Healy2024, Marasco2025, Veronese2025}, and create detailed kinematic models of extra-planar, diffuse \HI \citep{Kurapati2025, Zabel2026}. 

In this work, we identified star-forming galaxies of similar mass that are in close spatial and velocity proximity to each other in the MHONGOOSE survey \citep{deBlok2024}. Two fields satisfy this criteria, the NGC\,3511/3513 pair and the NGC\,4781 field, and we conducted a detailed analysis of the morphology and kinematics of the newly detected \HI down to \NHI = 10$^{17.9}$\,\cm. We also compared the distribution and kinematics of  the \HI gas to those of other gas components (molecular and ionised) and of stars to test whether mild environmental interactions end up disrupting the star-forming disc. 

This paper is organised as follows. In Sect. \ref{sec:sample}, we present the close galaxy systems in the MHONGOOSE survey. Section \ref{sec:Observations} describes the multi-wavelength data and their derived properties. The global properties of the galaxies, \HI distribution, scaling relations and kinematics of the star-forming disc are presented in Sect. \ref{sec:Results}. We discuss the origin of the newly detected \HI in Sect. \ref{sec:origin}. Our findings are summarised in Sect. \ref{sec:Summary}.

\section{Close galaxy systems in the MHONGOOSE survey}
\label{sec:sample}
\begin{figure*}
\centering
    \includegraphics[width = 0.93\textwidth]{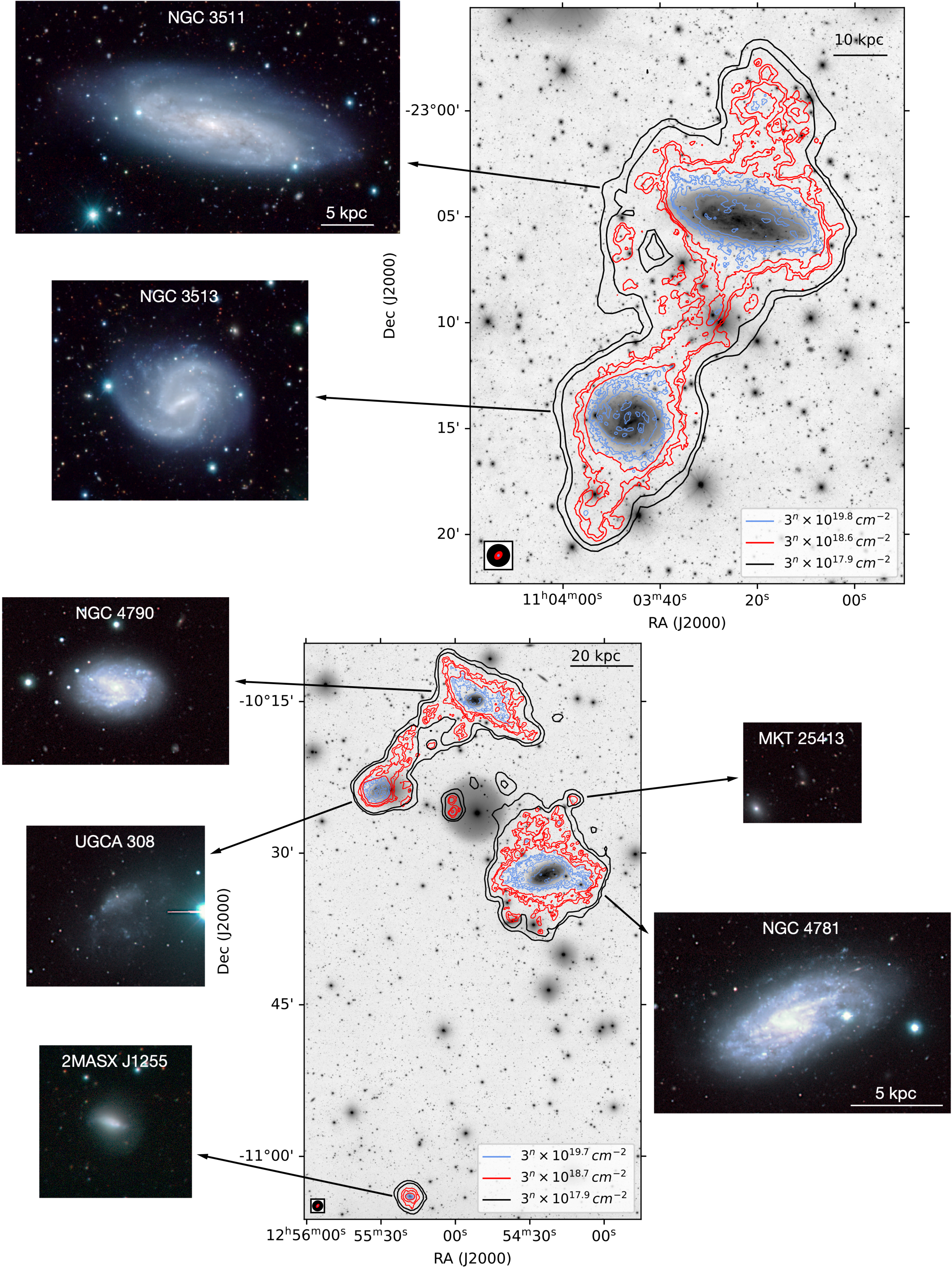}

\caption{Optical and \HI images for the close systems: the NGC\,3511/3513 pair (top) and the NGC\,4781 field (bottom). The optical RGB colour image of the individual galaxies shown as cutouts, and comprised of the VST $g$ and $r$-bands and DECaLS $z$-band. The grey-scale optical image is the VST $g$-band image, overlaid with primary beam-corrected \HI contours. The lowest \HI contour is the 3$\sigma$ value over 16\,\kms, where black, red and blue originate from the \texttt{r05\_t60}, \texttt{r10\_t00}, and \texttt{r00\_t00} images, respectively. To present the \HI structure and avoid over-crowding, we show n = 0, 1 for the \texttt{r05\_t60} and \texttt{r10\_t00} images, and n = 0, 1, 2, ... for the \texttt{r00\_t00} image. The synthesised beams of the different resolutions are shown in the bottom-left using the same contour colours, and a scale bar is shown in the top-right.}

\label{fig:overlay}
\end{figure*}

In this work, we aim to analyse the \HI distribution and properties of interacting star-forming galaxies of comparable mass. Therefore, we identified the MHONGOOSE fields containing at least two galaxies (including the target galaxy) with \HI masses within one dex of each other, a projected separation of $D_\mathrm{proj} \leq 100$\,kpc, and a systemic velocity difference of $\Delta v_\mathrm{sys} \leq 200$\,\kms. The two fields that meet these criteria are the NGC\,3511/3513 pair and the NGC\,4781 field. Figure\,\ref{fig:overlay} presents the optical colour images with the detected \HI emission overlaid on the VST $g$-band image, and the global properties of the galaxies are listed in Table\,\ref{tab:global_properties}.

\begin{table*}
    \centering
    \caption{Global properties of galaxies in close systems.}
   \begin{tabular}{ccccccc}
        \hline
        \hline
        Galaxy  & Morphology & log(\Mstar/\Msol) & log(\MHI/\Msol) & log(\Mdyn/\Msol) & log(SFR/\Msol yr$^{-1}$) & \vsys \\
        (1)     & (2)        & (3)               & (4)             & (5) & (6) & (7)\\
             &               &                   &                 &     &     & (\kms) \\
        
        \hline
        NGC\,3511                & SA(s)c$^{a}$       & 9.9 & 9.6 & 10.7 & -0.10 & 1099 \\
        NGC\,3513                & SB(rs)c$^{a}$      & 9.7 & 9.0 & 9.7 & -0.70 & 1191 \\
        \hline
        NGC\,4781                & SB(s)dm pec$^{b}$  & 9.5 & 9.2 & 10.5 & -0.32 & 1256 \\
        NGC\,4790                & SA(s)bc$^{b}$      & 9.1 & 8.8 & 10.1 & -0.52 & 1354 \\
        UGCA\,308                & dIrr$^{c}$         & 7.9 & 8.4 & 9.1 & -2.12 & 1317 \\
        2MASX\,J12551835-1103580 & dIrr$^{c}$         & 7.2 & 7.6 & 9.0 & -2.35 & 1105 \\
        MKT\,25413-102444        & --         & 6.4 & 5.7 & 7.9 & $<$-5.80 & 1334\\
        \hline
    \end{tabular}
    \tablefoot{(1) Galaxy name. (2) Optical morphology as measured by $^{a}$ \citet{Eskridge2002}, $^{b}$ \citet{Buta2015}, $^{c}$ \citet{Paturel2003}. (3) Stellar mass from \citet{deBlok2024}. (4) \MHI measured from the \texttt{r05\_t60} total intensity map. (5) Dynamic mass, estimated using the circular velocity from the kinematic modelling of each galaxy, except 2MASX\,J1255 and MKT\,25413, where $w_{\mathrm 20}$ was used. (6) SFR measured using the linear MIR+FUV hybrid method. The SFR of MKT\,25413 is the 3$\sigma$ FUV measured upper limit. (7) Systemic \HI velocity, measured by minimising the iso-velocity contour along the minor axis, of the highest resolution velocity field.}

    \label{tab:global_properties}
\end{table*}

\subsection{The NGC\,3511/3513 pair}
NGC\,3511 is a massive, late-type star-forming galaxy with a highly inclined ($i$ = 74\degree) stellar disc, and hosts a diffuse component in the outer regions (Fig. \ref{fig:overlay}). NGC\,3513 is approximately face-on (stellar inclination $i$ = 39\degree), lower mass barred spiral \citep{Eskridge2002, Zhu2024}. This was the only known galaxy pair in the MHONGOOSE survey. Both galaxies lie on the SF main sequence \citep{Renzini2015}, were known \HI detections \citep{Koribalski2004b}, and have well-defined stellar structure. A distance of 13.9 $\pm$ 2.1\,Mpc was estimated to NGC\,3511 using the numerical action model \citep[NAM:][]{Shaya2017, Anand2021}, making the projected distance between the galaxy pair to be 44\,kpc. As the virial radius of NGC\,3511 is 216\,kpc \citep{deBlok2024}, it is plausible that the pair is gravitationally bound.

The pair has previously been presented in \citet{Eibensteiner2024}, who first showed a filament of \HI connecting the galaxies from the MHONGOOSE observations. This study presented the radial distribution of the ratio between the atomic-to-molecular gas surface densities, and found that NGC\,3511 is \HI (not molecular gas) dominated. The H$_{\rm 2}$ radial profile for NGC\,3511 has the highest scatter of the sample, although the shape was consistent with the other \HI-rich, star-forming galaxies. The \HI filament did not manifest in any other of their measured properties, such as the SF surface density or dynamical equilibrium pressure \citep{Eibensteiner2024}.   

Long-slit optical spectroscopy along the major and minor axes of both galaxies has revealed the presence of extra-planar diffuse ionised gas \citep{Zhu2024}. In NGC\,3511, the broad component exhibits some rotational lag and was proposed to exist in the form of an extra-planar, gaseous disc. In contrast, the extra-planar ionised gas of NGC\,3513 is thought to be due to localised outflows at smaller (e.g. $\leq$ 1\,kpc) scales. The circulation of warm ionised gas at the disc-halo interface is seen in both galaxies \citep{Zhu2024}. 

\subsection{The NGC\,4781 field}
The MHONGOOSE target NGC\,4781 is an intermediate mass, dusty, star-forming, late-type galaxy \citep{deVacouleurs1959, Ho2011, Buta2015}, with a redshift-independent distance estimate of 11.3 $\pm$ 1.2\,Mpc, determined using the tip of the red giant branch \citep{Anand2021}. Prior to the MHONGOOSE survey, the \HI emission in the region of NGC\,4781 was confused with multiple sources in the HIPASS catalogue \citep{Meyer2004, Zwaan2004, Doyle2005, Koribalski2004b}. The coarse beam of HIPASS hinted at slightly extended emission compared to a point source, although the spectrum did now show evidence of additional \HI sources. Thanks to the deep, resolved, wide-field MeerKAT images of NGC\,4781 in the MHONGOOSE survey \citep{deBlok2024}, three additional galaxies are unambiguously detected in \HI at approximately the same velocity. They are NGC\,4790, UGCA\,308 and 2MASX\,J12551835-1103580 (hereafter 2MASX\,J1255), which were initially reported in the survey paper \citep{deBlok2024}. There is a fourth, \HI detection co-spatial with a low-surface brightness dwarf galaxy, on the edge of the diffuse extended \HI of NGC\,4781. We name this galaxy MKT\,25413-102444 (hereafter MKT\,25413) and the global properties of these galaxies are listed in Table \ref{tab:global_properties}. In Appendix\,\ref{sec:appen_MKT} we discuss the evidence for this being \HI associated with the galaxy, rather than a coincidental projection effect.

NGC\,4781 is the most massive late-type galaxy in the field, with a stellar disc inclination of $i$ = 68\degree \citep{Bournaud2005} and resides on the main sequence of the stellar mass, cold gas and SF global scaling relations. PHANGS has mapped the H$_{2}$ emission of NGC\,4781 \citep{Lang2020, Leroy2021, Querejeta2021} and found that it is not a candidate for molecular gas outflow \cite{Stuber2021}. NGC\,4790 is slightly less massive and more face-on, with a stellar inclination of $i$ = 44\degree, and a low velocity dispersion of the ionised gas observed with VLT/MUSE \citep{denBrok2020}. Neither UGCA\,308 nor 2MASX\,J1255 have been extensively studied in the literature and this is the first time that MKT\,25413 is being analysed. The VST image (Fig.\,\ref{fig:overlay}) shows that UGCA\,308, 2MASX\,J1255 contain young stars, while having a disturbed stellar population. MKT\,25413 is diffuse in nature, with an elongated stellar morphology along the NE-SW axis (Fig\,\ref{fig:MKT_overlay}). The virial radius of NGC\,4781 is 183\,kpc \citep{deBlok2024}, and if they are all at the distance of NGC\,4781, the projected distances between NGC\,4781 and NGC\,4790, UGCA\,308, 2MASX\,J1255 and MKT\,25413 are 61, 61, 114 and 25\,kpc respectively. Therefore, it is plausible that these galaxies are all bound to the same halo.

\section{Observations and ancillary data}
\label{sec:Observations}

\subsection{MeerKAT radio data}
The observing strategy and full data reduction procedure of the MHONGOOSE survey is described in \citet{deBlok2024}. Compared to the original data reduction, an additional image plane continuum subtraction\footnote{\url https://github.com/laduma-dev/mowjsub} (imcontsub) has been included prior to making the final cubes and moment maps. This procedure first masks the \HI emission in the image cube, and a low-order polynomial is fitted to each pixel, along the line of sight. The polynomial is then subtracted from the image cube, which removes residual continuum emission that is not visible in individual channels, but picked up during the original \SoFiA \citep{Serra2015a, Westmeier2021} source finding. The improved image cubes have a lower root-mean square (rms) by a few percent compared to the pre-imcontsub cubes, and results in some additional low-level \HI-emission being included in the final moment maps. 

Six standard \HI cubes are created over a range of resolutions that is sensitive to different \HI column densities \citep[see Table 4 in][]{deBlok2024}. We provide interactive 3D renders of all six resolutions of our galaxies (see Sect. \ref{sec:HI_Results}), although in this work we make use of the \texttt{r00\_t00}, \texttt{r10\_t00} and \texttt{r05\_t60} cubes, where the first number represents the robust value (e.g. \texttt{r05} is robust = 0.5) and the second number represent the Gaussian taper applied in the $uv$ domain (e.g. \texttt{t10} is a 10\arcsec\xspace taper). These cubes correspond to angular resolutions between $\sim$ 8 and 65\arcsec\xspace (0.5 -- 4.4\,kpc at 13.9\,Mpc and 0.4 -- 3.5\,kpc at 11.3\,Mpc) and provide a good balance between spatial resolution and column density sensitivity. The 3$\sigma$ \NHI sensitives range between 10$^{19.8}$ and 10$^{17.9}$\,\cm over 16\,\kms. In the most sensitive cubes, we can detect a 3$\sigma$ log(\MHI/\Msol) = 5.5 and 5.2 for a point source with a 50\,\kms line width at the distance of NGC\,3511 and NGC\,4781, respectively. 

\subsection{Optical imaging}
There is an ongoing campaign to observe all MHONGOOSE galaxies in in the Sloan $g$, $r$, and $i$ bands using the VLT Survey Telescope \citep[VST;][]{Schipani2012}. VST is a 2.6\,m optical telescope located at Cerro Paranal, Chile, and is equipped with OmegaCAM \citep{Kuijken2002}, a 32 CCD imager with a 0.21\arcsec\xspace resolution and an impressive field of view (FoV) of $\approx$ 1 $\times$ 1\,deg$^{2}$. The VST FoV is an excellent match with the MeerKAT primary beam full-width half maximum $\sim$ 1\,deg FoV at the \HI frequency.

The observing campaign (runIDs: 110.25AJ, 112.266Y, 113.26YW, 114.27U8) aims to detect low surface brightness emission, identify satellite galaxies and characterise their stellar properties, which we aim to present for all MHONGOOSE fields in a future paper. At the time of writing, the VST observations are ongoing, however, we note that the $g$ and $r$-band observations of the NGC\,3511 and NGC\,4781 fields were completed in April 2023. 

The observations were reduced using the Astro-WISE system \citep{McFarland2013}, which has been used for other projects such as the Kilo Degree Survey \citep[KiDS;][]{VerdoesKleijn2013}, the Fornax Deep Survey \citep[FDS;][]{Iodice2016, Venhola2018, Peletier2020} and the VST Early-type Galaxy Survey \citep[VEGAS;][]{Capaccioli2015, Iodice2020}. The data reduction follows the standard procedure of applying instrumental corrections for each frame using an overscan correction, removing the bias, flat-fielding, an illumination correction, masking out bad pixels, and background subtraction. Then, an astrometric calibration is applied so the images have world coordinates, followed by a zero point calibration, achieved by observing standard stars each night. Finally, all the frames are co-added together to create a mosaic. The typical seeing for the $g$ and $r$-band observations were $\leq$ 1.1 and 1.2\arcsec\xspace. The resulting mosaics reach a limiting surface brightness sensitivity of $\mu_{g}$ = 29.0 and $\mu_{r}$ = 28.0\,mag\,arcsec$^{-2}$. 

\subsection{Molecular gas data}
\label{sec:H2_data}
The Physics at High Angular resolution in Nearby Galaxies (PHANGS) project uses multi-wavelength observations to study sub-kpc galaxy scale physical processes in nearby galaxies \citep{Leroy2021}. PHANGS-ALMA is the molecular gas campaign, that observed the CO(2--1) emission line in 90 nearby, massive galaxies that lie on or near the SF main sequence \citep{Leroy2021}. NGC\,3511 and NGC\,4781 are both part of the PHANGS-ALMA sample, and at their respective distances the 2\arcsec\xspace CO resolution is able to resolve $\sim$ 135 -- 110\,pc structures. We make use of the CO observations with a 7.5\arcsec beam that has a 2\arcsec\xspace pixel size, identical to the \texttt{r00\_t00} \HI maps. To do a pixel-by-pixel comparison of the \HI and CO emission, we reproject the CO cubes and moment maps onto the grid of the \texttt{r00\_t00} \HI maps, which degrades the \HI velocity resolution by a factor of 1.8.

To estimate the molecular gas surface density, we use \SigHII = $\alpha_{CO}$ I$_{CO}$ where $\alpha_{CO(2-1)}$ = 6.25\,\Msol\,pc$^{-2}$\,(K\,\kms)$^{-1}$ as measured in \citet{Sun2018} and I$_{CO}$ is the CO integrated intensity in K\,\kms. The $\alpha_{CO(2-1)}$ used here includes the correction for helium and other heavy metals and and corresponds to the standard Galactic conversion factor \citet{Bolatto2013}. 

\subsection{Global star formation data}
There are many ways to measure SF, and the accuracy of each method often depends on the mass and type of galaxy. For example, massive late types contain large amounts of dust, which obscures the energetic photons emitted from young stars. However, these photons also proportionally heat the dust at infrared (IR) wavelengths, making the mid-IR (MIR) flux the dominant SF tracer at high masses. On the other hand, low-mass galaxies contain little-to-no dust and an unobscured SF tracer (e.g. UV wavelengths) is preferred. 

We adopted the linear hybrid approach to measure SF rates (SFRs), as described in \citet{Leroy2019}. This combines the SF measured in MIR ($\lambda$ $\approx$ 22\,$\mu$m) from the Wide-field Infrared Explorer \citep[WISE;][]{Wright2010, Jarrett2013} W4 band with the SF measured in the far-ultraviolet (FUV; $\lambda$ $\approx$ 154\,nm) from the Galaxy Evolution Explorer \citep[GALEX;][]{Martin2005}. The hybrid method accounts for the obscured and unobscured SF contributions. It works well for star-forming galaxies \citep[see][]{Leroy2019} and it is more robust than measuring SFRs in a single band. 

The SFRs measured here differ slightly from those presented in the survey paper by \citep{deBlok2024}, where a WISE MIR SFR was scaled with a UV flux correction \citep{Cluver2025}. The latter method is adequate for large statistical samples, but introduces substantial uncertainty in the UV contribution for individual (especially low mass) galaxies \citep[for details, see][]{Cluver2025}. As we only had a handful of galaxies at our disposal, with most detected in both WISE MIR and GALEX FUV, we opted for the hybrid method for measuring global SFRs.

\subsection{Spatially-resolved stellar and ionised gas observations}
NGC\,3513 and NGC\,4790 were observed with the Multi Unit Spectroscopic Explorer (MUSE) instrument in wide-field mode nominal wavelength range under program IDs 098.B-0551 and 097.B-0165 respectively. In both cases, the observations consisted of 4 $\times$ 900\,s frames on target and a 3 $\times$ 120\,s offset sky exposure taken immediately after. Data were reduced by the ESO pipeline team using the ESO Recipe Execution Tool, version 3.12.3 \citep{esorex2015}, and MUSE Instrument Pipeline v1.6.4 \citep{Weilbacher2020}, while the fully reduced cube was downloaded from the ESO science archive.

The reduced cubes for NGC 3513 and NGC 4790 were analysed using the \textsc{nGIST} pipeline\footnote{\href{https://ascl.net/2507.015}{https://ascl.net/2507.015}}, as described in \citet{Fraser-McKelvie2025}. Stellar and ionised gas kinematics were derived using \textsc{nGIST}'s ppxf routine for the \texttt{KIN} and \texttt{GAS} modules, respectively. In each case, the MUSE cubes were Voronoi binned to a continuum signal-to-noise ratio of 50, and the cubes analysed over the wavelength range of 4900--7000 \AA. For the gas kinematics derived via bright emission line measures, we also produce data products on a spaxel scale ($0.2^{\prime\prime} \times 0.2^{\prime\prime}$). \textsc{nGIST} outputs 2D maps of light-weighted mean stellar line of sight velocity ($v_{\star}$), velocity dispersion ($\sigma_{\star}$), and ionised gas velocity as traced by H$\alpha$, $v_{H\alpha}$, and ionised gas velocity dispersion, $\sigma_{H\alpha}$. 

\section{Results}
\label{sec:Results}

\subsection{Global Properties}
\begin{figure}
\centering
    \includegraphics[width = \columnwidth]{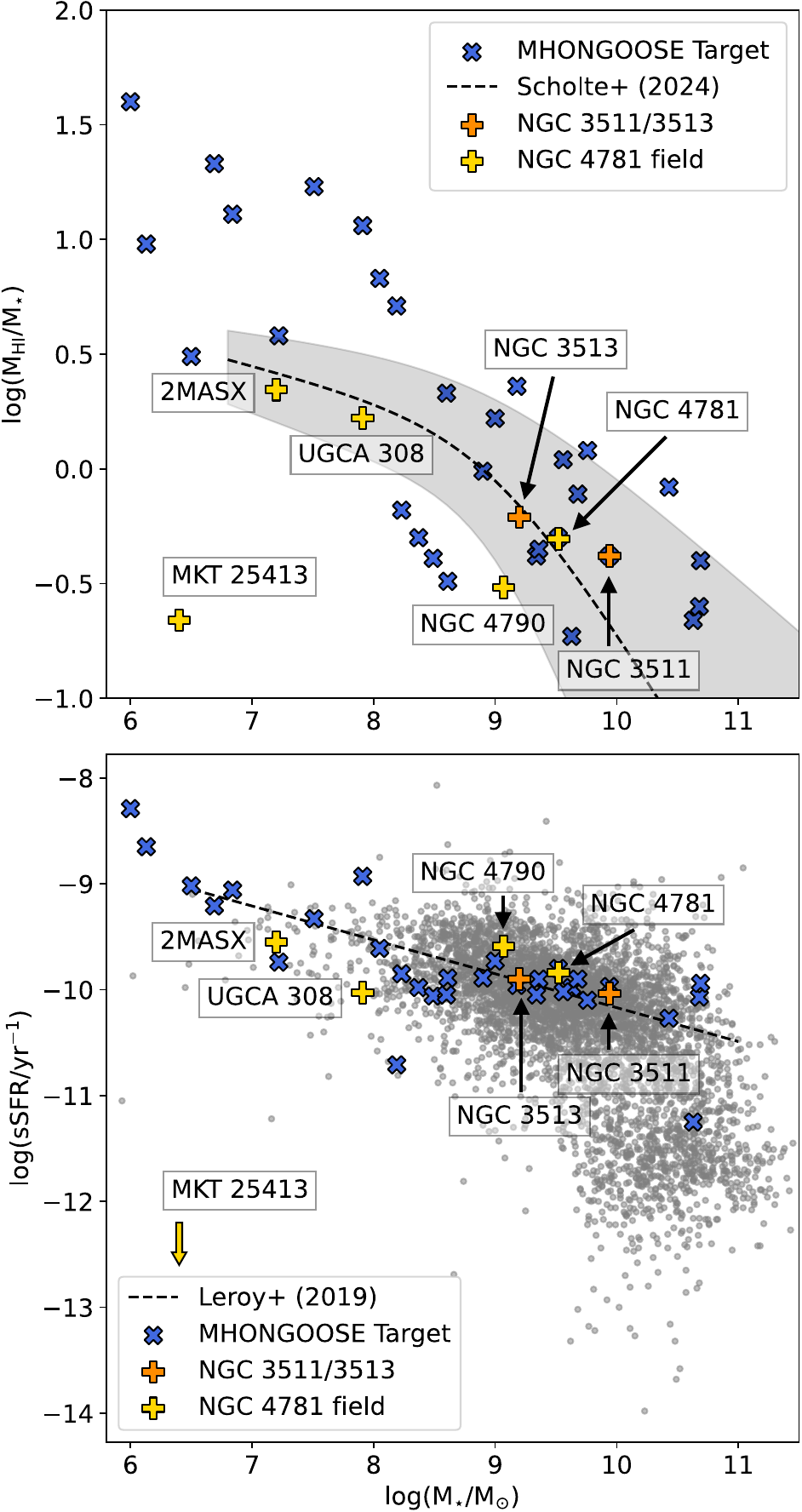}

\caption{Global scaling relations. \textit{Top}: \HI-to-stellar mass (\FHI) ratio as a function of stellar mass. \textit{Bottom}: SFR-to-stellar mass (sSFR) ratio as a function of stellar mass. The blue crosses are all the MHONGOOSE targets, the orange plus markers denote the NGC\,3513/3511 pair while the yellow plus markers are galaxies from the NGC\,4781 field. The dashed black line in the top panel is the \citet{Scholte2024} median \FHI relation with the grey shaded region representing the uncertainty in the parametric fit. The grey points in the bottom panel are galaxies from \citet{Leroy2019} with the sSFR main sequence shown as the dashed black  line. Six of the seven \HI-detected galaxies in this work lie on the main \FHI and sSFR sequences and are comparable to other gas-rich star-forming galaxies. MKT\,25413 is an outlier and lies substantially both scaling relations.}

\label{fig:scaling}
\end{figure}

We place all seven \HI-detected galaxies on two global scaling relations: the \HI gas fraction of \FHI $\equiv$ log(\MHI/\Mstar) and the specific SFR of sSFR $\equiv$ log(SFR/\Mstar), as a function of stellar mass (Fig.\,\ref{fig:scaling}). Six of our \HI-detected galaxies lie within the scatter of the \citet{Scholte2024} \FHI scaling relation (Fig.\,\ref{fig:scaling}, upper panel). The exception is MKT\,25413, for which we measure log(\MHI/\Msol) = 5.7, the lowest \HI mass measurement\footnote{MKT\,25413 has the second lowest \MHI of all galaxies detected in the MHONGOOSE survey. The lowest mass belongs to a galaxy in the J0310--39 field with log(\MHI/\Msol) = 5.5 at a distance of 8\,Mpc, and will be presented in future work.}, that is not an upper limit, in the literature for a galaxy that is not a Milky Way satellite. It lies well below the relation by $\sim$\,1 dex, making it very \HI-poor. We show all MHONGOOSE target galaxies (blue crosses) in Fig.\,\ref{fig:scaling} and note that those with log(\Mstar/\Msol) $<$ 8 systematically reside above the \FHI scaling relation. This is due to MHONGOOSE targeting gas-rich, star-forming galaxies in low-density environments, unaffected by strong environmental effects, whereas \citet{Scholte2024} used a stellar-mass selection that is not biased towards gas-rich galaxies.

We converted the SFR-\Mstar relation of \citet{Leroy2019} to sSFR-\Mstar and extrapolated down to lower masses as a visual guide for the lower mass MHONGOOSE galaxies (Fig.\,\ref{fig:scaling}, lower panel). As first reported in \citet{deBlok2024}, the MHONGOOSE targets (with few exceptions) lie on the sSFR main sequence. The same six galaxies with typical \HI-to-stellar mass ratios also lie on the sSFR main sequence (Fig.\,\ref{fig:scaling}). UGCA\,308 is slightly below the relation, although within the scatter of star-forming galaxies at that stellar mass. It is only MKT\,25413 that has no measurable SF in the GALEX near-UV (NUV) and FUV images, nor in the SINGG \citep{Meurer2006} \Ha image. The 3$\sigma$ SF upper limit measured in FUV is shown in (Fig.\,\ref{fig:scaling}) and places it at least 3\,dex below the sSFR main sequence at that stellar mass. 

The global properties for NGC\,3511, NGC\,3513, NGC\,4781, NGC\,4790, UGCA\,308, and 2MASX\,J1255 are comparable to other star-forming, \HI-rich late-type galaxies of the same stellar mass. As described above, they lie on the \FHI and sSFR main sequences, display a well-defined spiral structure, and contain dust. UGCA\,308 and 2MASX\,J1255 have irregular optical morphologies, which is common for galaxies below log(\Mstar/\Msol) = 8. Only MKT\,25413 lies substantially below the \FHI and sSFR main sequences and appears as a low-surface brightness galaxy with a small amount of \HI and no SF. 

\subsection{\HI emission, distribution and kinematics}
\label{sec:HI_Results}
The \HI emission is extremely resolved and a wealth of new features are present at both low-and-high column density. We present the total intensity (moment zero) from three cubes at three different resolutions and the \texttt{r10\_t00} position-velocity (PV) diagrams of the interacting galaxies along their kinematic major axis, determined by 3D tilted ring fitting (detailed in Sect.\,\ref{sec:disk}), for the close systems in Fig.\,\ref{fig:mom0}. We present the resolved \HI emission in 2D and 3D in Appendix\,\ref{sec:appen_dist} in the form of channel maps and interactive 3D renderings \citep{Labadie-Garcia2025}. We highly encourage the reader to utilise and explore the 3D renders as it intuitively shows the structures and morphology of the \HI emission better than a 2D projection. To verify the robustness of the described features, we inspect multiple cubes at different resolutions in virtual reality (VR) using the \texttt{iDaVIE-v} software \citep{Marchetti2022, Sivitilli2026}. This ensures that tails, bridges and \HI associated with each galaxy are not projection effects.

\begin{figure*}
\centering
    \includegraphics[width = 0.95\textwidth]{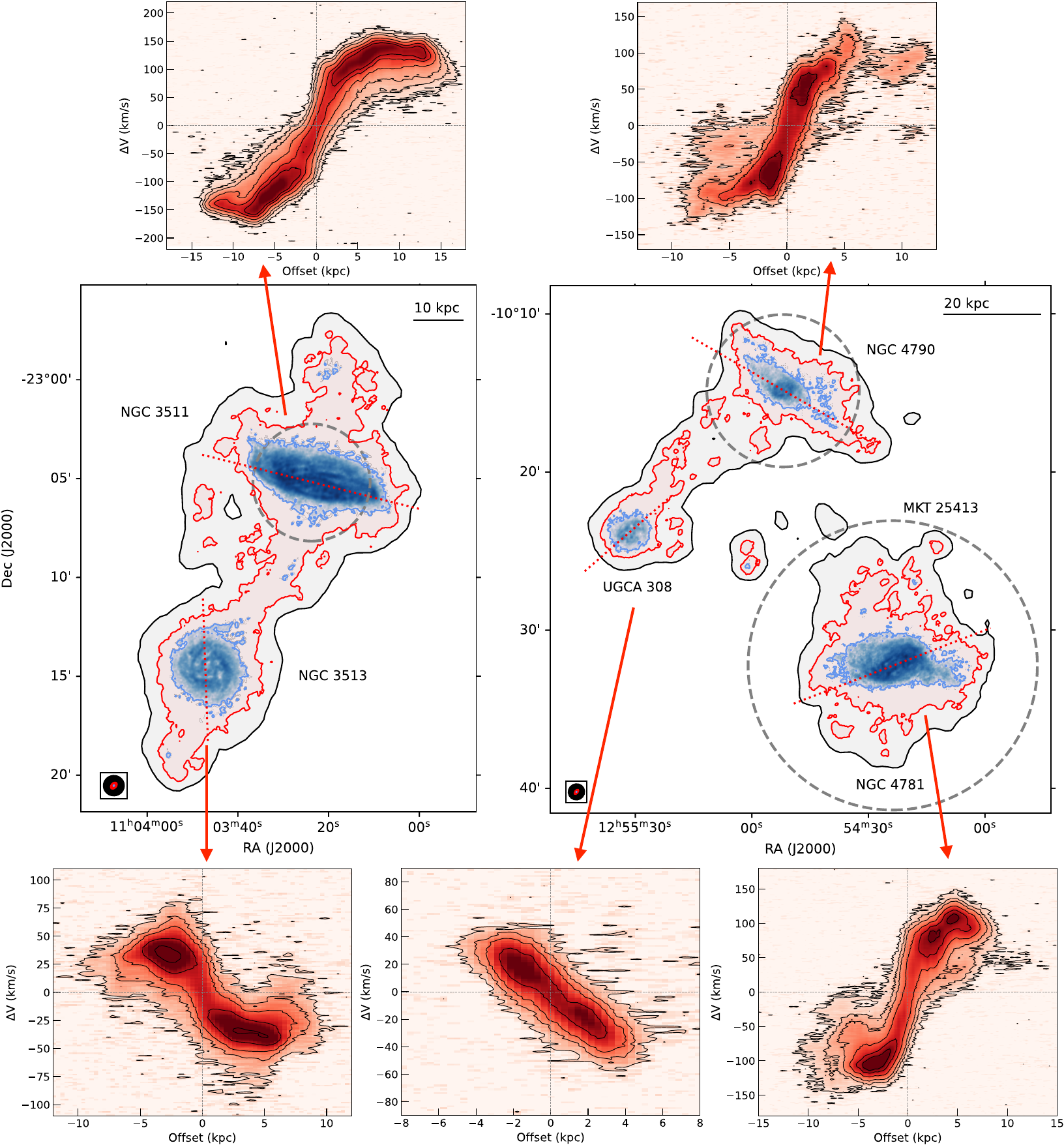}

\caption{Close systems total \HI intensity and PV maps, together with the moment-zero map of the NGC\,3511/3513 pair (\textit{left}) and the interacting galaxies in the NGC\,4781 field (\textit{right}). The contour colours correspond to the same datasets as in Fig. \ref{fig:overlay}. The contour for each resolution is the 3$\sigma$ value over 16\,\kms, the same as Fig.\,\ref{fig:overlay} for n = 0. A scale bar is situated in the top-right, and the synthesised beams of each resolutions are shown in the bottom-left using the same colours as their respective moment zero image. The dashed grey circles represent the tidal radius of each galaxy, computed with respect to the next lower mass galaxy in the system. The red arrow points to the \texttt{r10\_t00} position-velocity diagram of the respective galaxy, taken along the kinematic major axis (red dotted line in the moment zero map). The contours in the PV diagrams are at 3, 10, 20, 50, 100, and 150$\sigma$, and show an abundance of low column density, extra-planar \HI.} 

\label{fig:mom0}
\end{figure*}

The \HI emission in Fig.\,\ref{fig:mom0}, reveals that NGC\,3511 and NGC\,3513 are currently interacting. The dense \HI above \NHI = 10$^{19.8}$\,\cm (the blue contours in the Figs.\,\ref{fig:overlay} and \ref{fig:mom0}) traces the stellar structures (i.e. disc and spiral arms). Despite the interaction, there are almost no \HI clouds at this density outside the stellar discs. At \NHI = 10$^{18.6}$\,\cm (the red contours in Figs.\,\ref{fig:overlay} and \ref{fig:mom0}) there is a 44\,kpc projected length \HI bridge connecting the two galaxies, extended \HI emission pointing to the north for NGC\,3511 and the south for NGC\,3513, and a series of \HI clouds to the east of NGC\,3511 and the bridge. The pair is embedded in a diffuse \HI envelope with a column density of \NHI = 10$^{17.9}$\,\cm (the black contours in Figs.\,\ref{fig:overlay} and \ref{fig:mom0}).

In the NGC\,4781 field, the \HI emission (Fig.\,\ref{fig:mom0}) reveals a complex, interacting triplet between NGC\,4781, NGC\,4790 and UGCA\,308. The dense \HI above \NHI = 10$^{19.8}$\,\cm (the blue contours in the Figs.\,\ref{fig:overlay} and \ref{fig:mom0}) of NGC\,4781 extends beyond the stellar disc in the E-W direction and is misaligned by $\sim$ 30\degree with the optical major axis. The \HI at this column density extends in the NE-SW direction for NGC\,4790. At \NHI = 10$^{18.6}$\,\cm (the red contours in Figs.\,\ref{fig:overlay} and \ref{fig:mom0}) there is \HI emission above and below the discs of NGC\,4781 and NGC\,4790, which can also be seen in the PV diagrams, and UGCA\,308 hosts a long \HI tail pointing directly towards NGC\,4790. At this column density, the \HI of MKT\,25413 is detected along with two clouds in the IGM. The diffuse \HI at \NHI = 10$^{17.9}$\,\cm (the black contours in Figs.\,\ref{fig:overlay} and \ref{fig:mom0}) connects NGC\,4790 and UGCA\,308 with a bridge, extends the \HI above and below the disc of NGC\,4781 and we detect two additional clouds in the IGM. Unlike the interacting galaxies, the \HI emission of 2MASX\,J1255 is symmetric over the full column density range. Here, we focus the analysis on the interacting triplet in the field, with a discussion of 2MASX\,J1255 and MKT\,25413 presented later in this work.

\begin{figure*}
\centering
    \includegraphics[width = \textwidth]{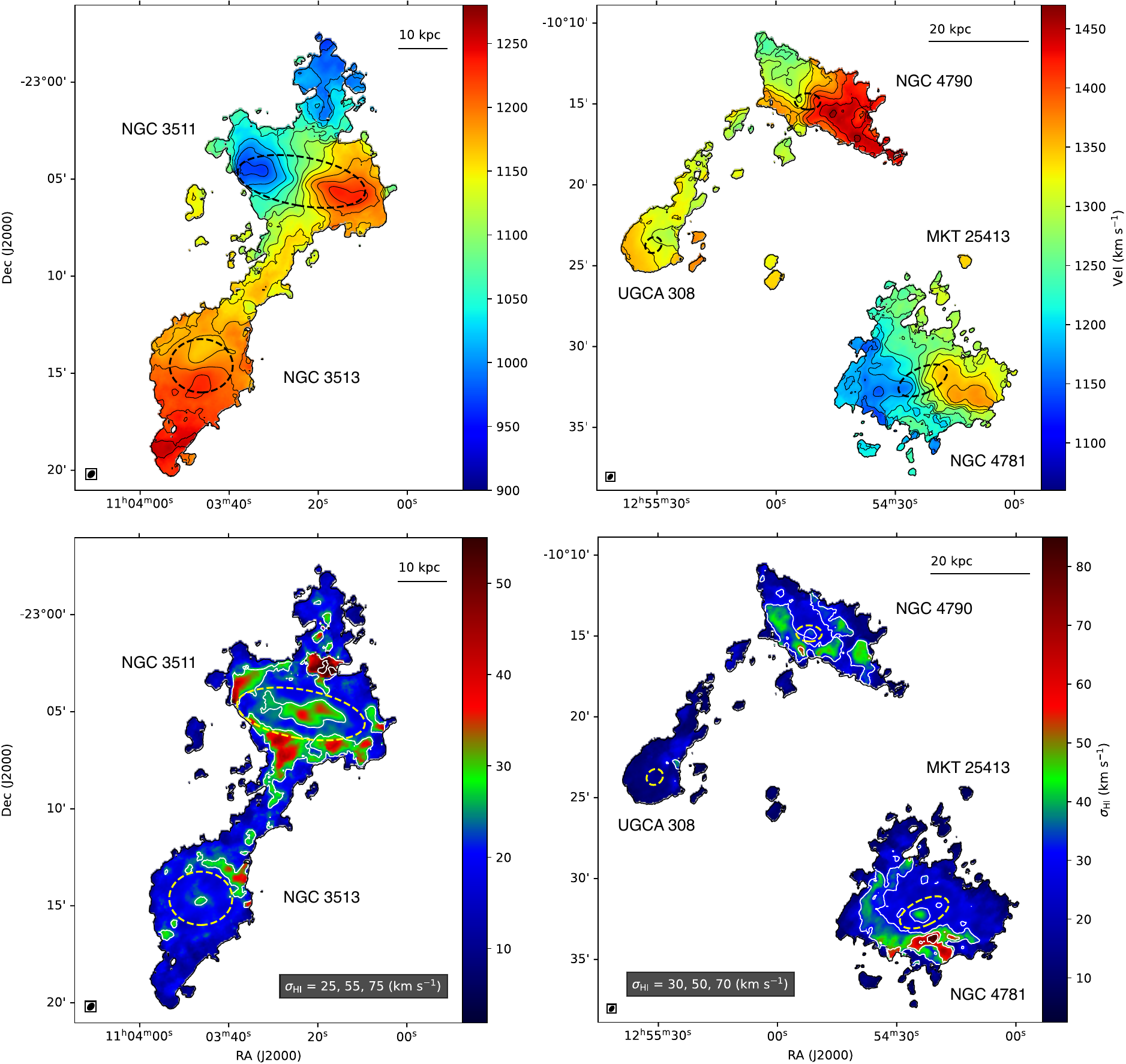}

\caption{\HI velocity and velocity dispersion fields. \textit{Top}: \texttt{r10\_t00} velocity field (moment one) of the close systems. \textit{Bottom}: Line-of-sight velocity width (moment two). \textit{Left}: NGC\,3513/3511 pair. \textit{Right}: Interacting galaxies in the NGC\,4781 field. The velocity field is centred on the systemic velocity of the MHONGOOSE targets (NGC\,3511 and NGC\,4781), with contours in steps of 25\,\kms. The dashed ellipses (black in the moment one and yellow in the moment two maps) denote the extent of the stellar discs. The beam ($\sim$ 28\arcsec $\times$ 17\arcsec) is shown in the bottom-left, and a scale bar on the top-right. High moment two values are highlighted by the white contours with their values displayed in the inset. The velocity field shows rotation of the inner-discs, coherent structures and many kinematically anomalous components. The high moment two values indicate coherent \HI, wrapping around the stellar discs.}

\label{fig:vel_field}
\end{figure*}

We show the \texttt{r10\_t00} \HI velocity field (moment one) and line-of-sight velocity width (moment two) maps of the close systems in Fig.\,\ref{fig:vel_field}. The velocity field shows there are a number of kinematically distinct features. First, the \HI of all the inclined late-type galaxies (NGC\,3511, NGC\,4781 and NGC\,4790) show very regular and symmetric rotation within the stellar disc (dashed ellipses in Fig.\,\ref{fig:vel_field}). Beyond the stellar disc, the velocity structure of NGC\,4781 is heterogenous with many asymmetries, indicating more complex gas dynamics than a rotating disc. It can be seen in NGC\,4790 that the kinematic major axis of the \HI beyond the stellar disc is offset from the rotation within the stellar disc (also seen in the PV diagram Fig.\,\ref{fig:mom0}). The \HI in the main bodies of \HI of NGC\,3513 and UGCA\,308 show a smooth velocity gradient (also seen in the PV diagram in Fig.\,\ref{fig:mom0}) that indicate rotation. The extension to the south of NGC\,3513 and the main component both appear to be smooth extensions of the galaxy. Similarly for UGCA\,308, the velocity structure of the tail smoothly connects to the galaxy. The velocities of the extended emission to the north of NGC\,3511 is offset by $\sim$200\,\kms, and therefore a kinematically distinct component to the rotating disc. The clouds to the east of NGC\,3511 comprise the dense component of an arc structure, and have a smooth velocity gradient connecting to the bridge. The velocity differs from the galaxy rotation velocity on the same side by $\sim$200\,\kms, though the clouds may be a continuation of the extended emission to the north of NGC\,3511.

The inclined late-type galaxies (NGC\,3511, NGC\,4781 and NGC\,4790) exhibit coherent, large-scale arcs of high moment-2 values $\sigma_\mathrm{HI}$ $>$ 30\,\kms encircling the stellar disc (Fig.\,\ref{fig:vel_field}). There is a potential coherent arc outside wrapping around the stellar disc of NGC\,3513, although shorter than the other arcs. These arcs are diffuse, with \NHI $\sim$ 10$^{17.9}$\,\cm and can be described as bulk flows of \HI gas wrapping around the stellar disc. Given the nominal \HI line broadening of $\sim$15\,\kms, the typical moment two values (30 -- 45\,\kms) in the arcs are significantly higher, with peaks reaching up to 80\,\kms. The \HI spectra within these regions show a variety of features. In many instances, individual profiles host multiple peaks indicating that the gas flow is clumpy and the high moment-2 values are not purely a result of thermal turbulent broadening. Where a bridge is present, the arcs are closest to the bridge that connects the galaxies. In the case of NGC\,4781, one side of the arc points to the debris in the north. Therefore, the multiple components responsible for the increased line width may be a superposition of the rotating disc, debris, and the bridge (when present).

\subsection{The properties of the disc and multiphase ISM}
\label{sec:disk}
In this section, we focus on the multi-wavelength observations that trace the different (cold molecular gas, ionised gas and stellar) components of the disc and ISM. We compare the different phases to the \HI, and where possible, we quantify the kinematics through the use of the tilted-ring fitting software \BBarolo \citep{diTeodoro2015}. As our aim is to probe the star-forming disc, we use measurements of the kinematics at the highest resolution (\texttt{r00\_t00} -- 2\arcsec\xspace pixels, 8\arcsec\xspace beam) \HI cube. Following the standard two-step procedure used in \citet{diTeodoro2021}, we first fit the geometric parameters, such as the centroid, inclination, and position angles. Then, the geometric parameters were fixed and we fit for the rotation velocity and dispersion. Both the approaching and receding sides were fitted, although the fit was axisymmetric during the fitting process for each side. 

\subsubsection{Molecular gas distribution and kinematics}
The central galaxies of the close systems (NGC\,3511 and NGC\,4781) were observed as part of the PHANGS–ALMA project \citep{Leroy2021}. The CO maps are limited to the inner $\sim$ 50 and 70\% of the stellar disc for NGC\,3511 and NGC\,4781, respectively. The CO kinematics is fitted using the same two-step procedure as the \HI, although the fit was done independently and only the \HI geometry was used as an initial guess in step one of the fitting. 

\begin{figure}
\centering

    \includegraphics[width = 0.75\columnwidth]{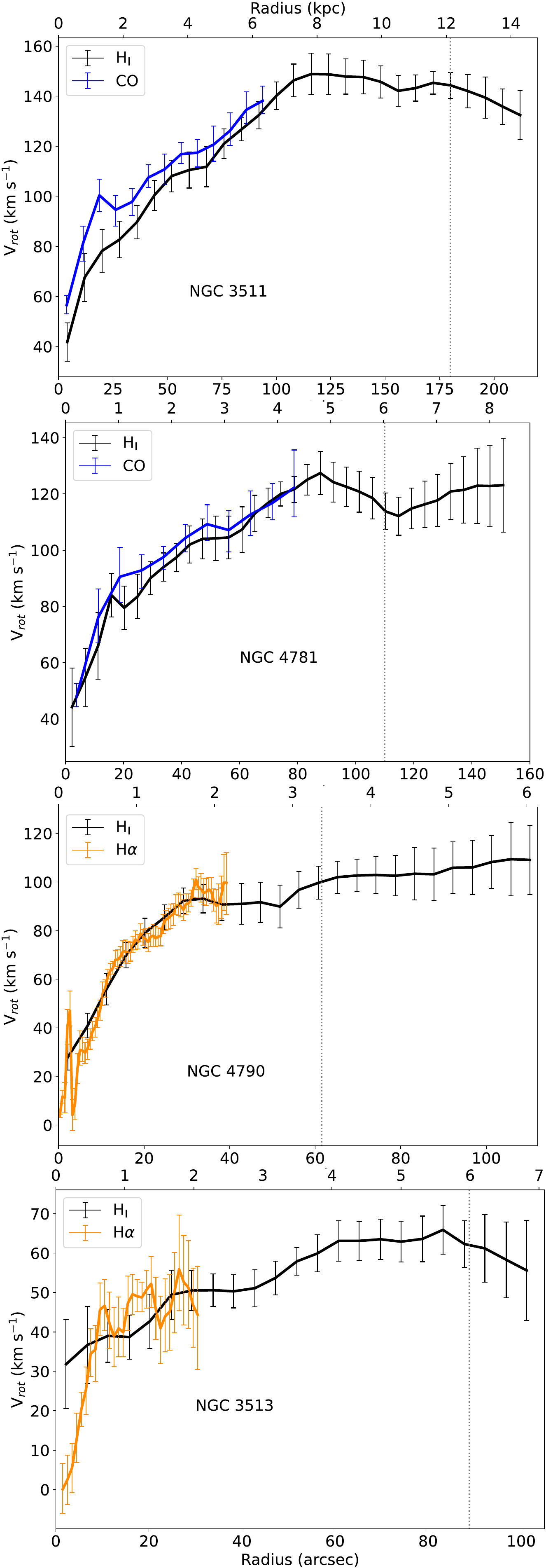}

\caption{Rotation curves. The tilted ring fitted \HI (black), CO (blue) and \Ha (orange) rotation curves, where the dotted grey line represents the extent of the stellar disc. The bottom axis shows the angular radius and the top is the physical radius. The rotation curves of the different gas phases closely follow each other and imply co-rotation. There is a slight lag between the \HI and CO of NGC\,3511.}

\label{fig:rot_curves}
\end{figure}

In Fig.\,\ref{fig:rot_curves}, we present the \HI and CO rotation curves for NGC\,3511 and NGC\,4781. For both galaxies, the CO rotation curve closely follows the \HI. The CO of NGC\,4781 just reaches the flat part of the rotation curve (v$_{rot}$ = 120\,\kms) whereas the CO only samples the rising part of the rotation curve for NGC\,3511. The very inner part of NGC\,3511 may rise slightly steeper in CO compared to the \HI, indicating a slight lag in the rotation of the atomic hydrogen. This effect may tentatively also be seen in NGC\,4781, although it is less pronounced and within the uncertainties of the tilted-ring fitting.

\begin{figure*}
\centering
    \includegraphics[width = \textwidth]{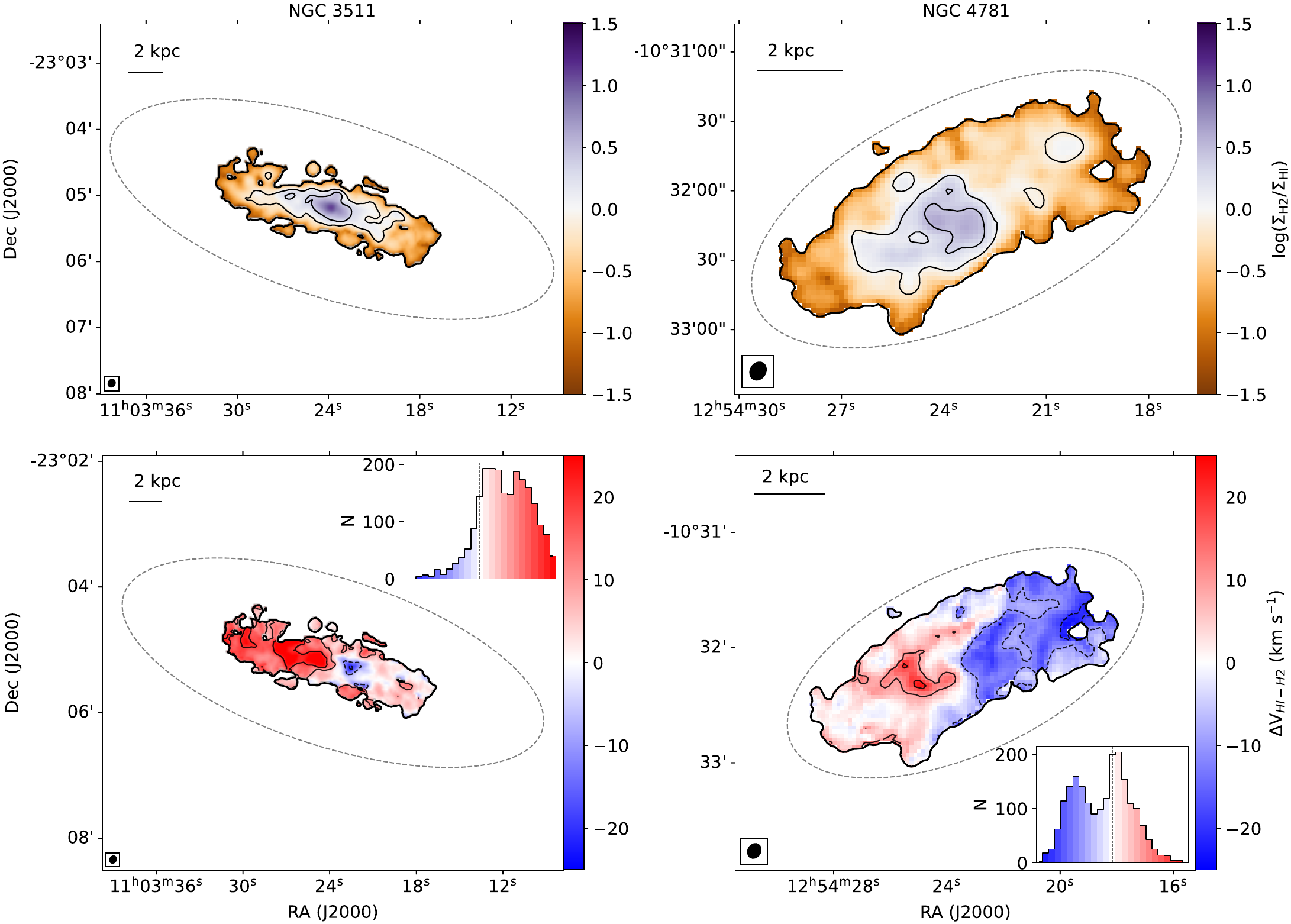}

\caption{Molecular-to-atomic hydrogen maps for NGC\,3511 (left column) and  NGC\,4781 (right column). \textit{Top panels}: Molecular-to-atomic hydrogen surface density ratio. The dashed grey ellipse represents the extent of the stellar disc, a scale-bar is placed in the top-left and the synthesised beam in the bottom-left. The black contours are molecular hydrogen surface density values of \SigHII = 1, 10 and 50\,\Msol\,pc$^{-2}$. \textit{Bottom panels}: Difference between the \HI and H$_{2}$ velocity fields. The dashed grey ellipse, scale-bar, and synthesised beam are the same as the top panel. \textit{Inset}: $\Delta{V}_{HI - CO}$ pixel histogram, matched to the colour-bar. The $\Delta{V}_{HI - H2}$ show only small differences between the kinematics, and the slight  NGC\,3511 \HI lag can be seen in the east.}

\label{fig:Rmol_maps}
\end{figure*}

As mentioned in Sect. \ref{sec:H2_data}, we used the $\alpha_{CO(2-1)}$ conversion factor from \citet{Sun2018} to estimate the H$_{\mathrm 2}$ mass from the CO. The molecular-to-atomic hydrogen surface density ratio ($\Sigma_{\mathrm{ H2}}$/$\Sigma_{\mathrm{HI}}$) and the difference between the \HI and CO velocity fields are shown in Fig.\,\ref{fig:Rmol_maps}. Where both \HI and CO are detected, the highest molecular-to-atomic ratios (up to $\sim$ 10) are only found in the core of NGC\,3511, where \SigHII $\gtrsim$ 50\,\Msol\,pc$^{-2}$ (Fig.\,\ref{fig:Rmol_maps}). NGC\,4781 shows a clear east–west asymmetry: eastern complexes reach $\Sigma_{\mathrm{ H2}}$/$\Sigma_{\mathrm{HI}}$ $\sim$ 10, with \SigHII $\approx$ 50\,\Msol\,pc$^{-2}$, whereas the western side is more uniform with $\Sigma_{\mathrm{H2}}$/$\Sigma_{\mathrm{HI}}$ $\sim$ 1, and \SigHII $\sim$ 1\,\Msol\,pc$^{-2}$. There is no high surface-density \HI coincident with the eastern molecular gas excess. The stellar disc is brighter on the eastern side (Fig.\,\ref{fig:overlay}), and the same east-west asymmetry is present in the WISE W3 and W4 bands, which trace dust and is indicative of UV being absorbed from massive SF regions.

The CO velocity fields of both galaxies display regular rotation, and the iso-velocity contours show no evidence of a warp. There are minor asymmetries between the approaching and receding sides, but they are insufficient to indicate substantial departures from ordered rotation. The differences between the \HI and CO velocity fields are low. The majority of the $\Delta{V}_{HI - CO}$ distribution is less than 10\,\kms, where the tails of the distribution reach a maximum 25\,\kms (Fig.\,\ref{fig:Rmol_maps}), which is smaller than $\sim$ 2$\times$ the intrinsic \HI broadening. The slight lag of \HI rotation can be seen in the east of NGC\,3511 and (to a lesser extent) the west of NGC\,4781 in the $\Delta{V}_{HI - CO}$ map (Fig.\,\ref{fig:Rmol_maps}), where the median difference is 9 and 2\,\kms respectively. As CO (and thus H$_{2}$) is dynamically colder and less pressure-supported than the \HI, this is likely an effect of asymmetric drift that typically results in a difference of $\sim$ 5-10\,\kms between the atomic and molecular gas phases, consistent with our values. 

\subsubsection{Ionised gas and stellar kinematics}
The two late-type satellites (NGC\,3513 and NGC\,4790) in the close systems have been observed with MUSE, allowing us to quantify the ionised gas and stellar kinematics. The MUSE (1\arcmin $\times$ 1\arcmin) FOV covers the central region of NGC\,3513, which contains the bar and inner spiral arms, whereas most of the stellar disc is covered for NGC\,4790. From the \texttt{nGIST} pipeline, native resolution spaxel and binned maps with a signal-to-noise ratio of S/N = 50 are created. As these galaxies are not in the massive regime, their rotation velocities are low, such that the \Ha emission is not contaminated by N\,\textsc{ii}\xspace emission in the channels of the MUSE cubes. Therefore, we run \BBarolo on the \Ha emission to kinematically model the ionised gas component of the ISM. Due to the coarse channel resolution of MUSE, the flux of the galaxies are restricted to fewer (five in the case NGC\,3513) channels compared to the \HI. Despite this limitation, it is enough to produce a simplistic kinematic model, and rather than a two-stage fit, we only fit for the rotation velocity, velocity dispersion and position angle and fix the geometric parameters, such as the centroid and inclination.

In Fig.\,\ref{fig:rot_curves}, we show the rotation curves for the \HI and \Ha of NGC\,3513 and NGC\,4790. The \Ha rotation curve of NGC\,4790 steeply rises and just reaches the flat part where v$_{rot}$ = 100\,\kms, which matches that of the \HI (Fig.\,\ref{fig:rot_curves}). We assumed a constant inclination of $i$=65\degree, although we tested different inclinations that range between 55 and 75\degree (based of the \HI kinematic fitting in the inner two kpc) and found that the shift of the \Ha rotation curve was less than the \BBarolo uncertainties. In the case of NGC\,4790, the \Ha and \HI rotation curves broadly agree, although there is a mismatch in the very inner part, where the \Ha is rising more steeply that the \HI (Fig.\,\ref{fig:rot_curves}). This is a resolution effect of the coarser spatial \HI resolution, which assigns the most inner ring a rotation velocity of $\sim$30\,\kms. The similarities in rotation curves for both galaxies imply that the \Ha and \HI are co-rotating as part of the same disc.

\begin{figure*}
    \centering
    \includegraphics[width = \textwidth]{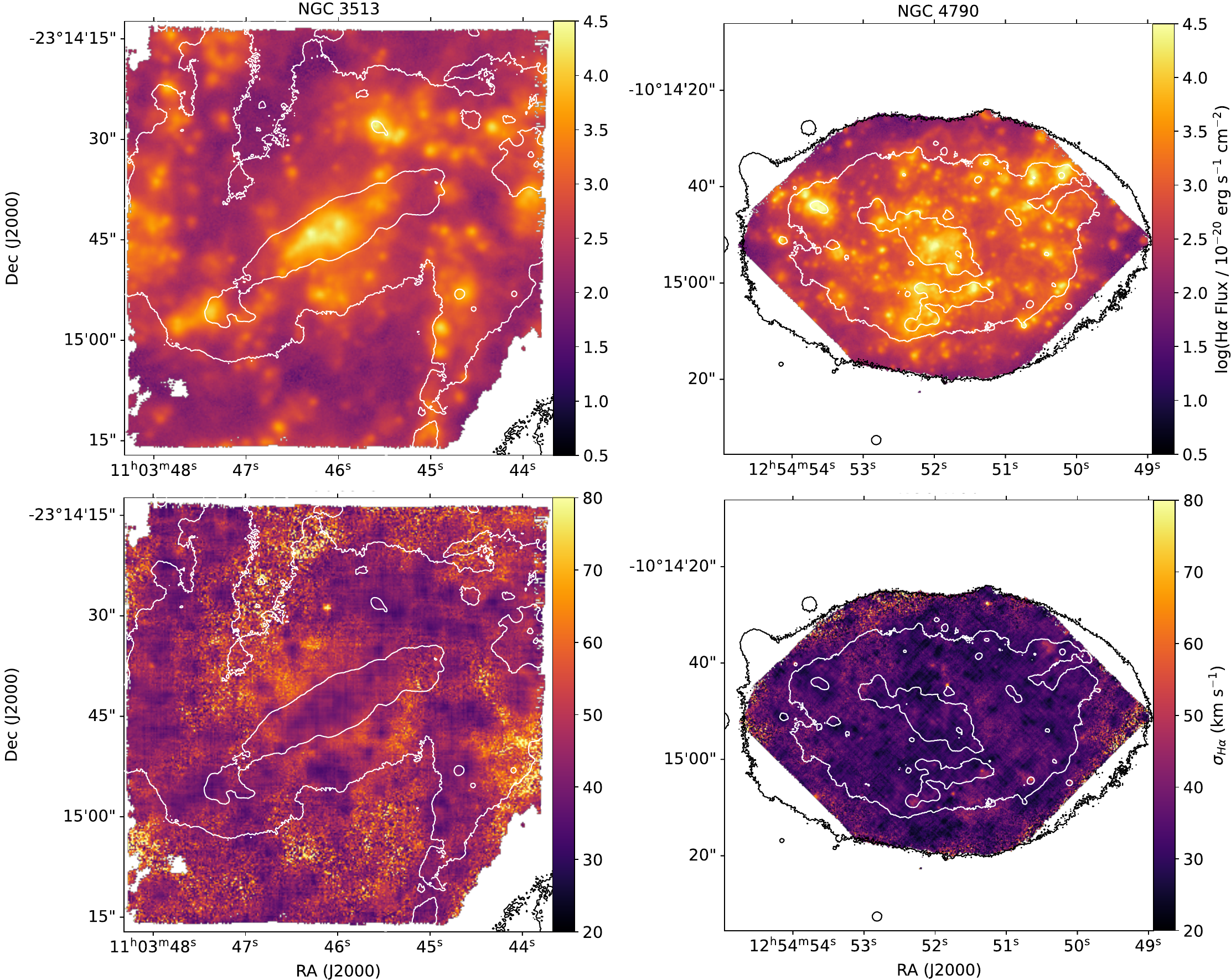}

\caption{Ionised gas maps for NGC\,3513 (left) and NGC\,4790 (right).\textit{ Top}: Spaxel \Ha flux map. The black and white contours trace $\mu_{g}$ = 23.25, 21.75,2 and 20.75\,mag\,arcsec$^{-2}$. \textit{Bottom}: Spaxel \Ha velocity dispersion with the same surface brightness contours. The regions of high \Ha flux correspond to regions of low $\sigma_{H\alpha}$.}

\label{fig:MUSE_Ha_maps}
\end{figure*}

The \Ha flux spaxel and velocity dispersion maps are presented in Fig.\,\ref{fig:MUSE_Ha_maps}. In both galaxies, the \Ha emission is clumpy and traces the optical structures, such as the bar and spiral arms outlined by the high surface brightness contours (Fig.\,\ref{fig:MUSE_Ha_maps}). As \Ha typically traces SF, the highest SF region in NGC\,3513 is in the bar, whereas NGC\,4790 hosts many star-forming knots throughout the disc, responsible for its slightly elevated sSFR (Fig.\,\ref{fig:scaling}). We see prominent bar effects in the NGC\,3513 \Ha velocity dispersion map, where high values ($\sigma_{H\alpha}$ $>$ 75\,\kms) are offset from the bar (Fig.\,\ref{fig:MUSE_Ha_maps}). For example, it is possible that complex kinematics around the bar are causing broader line profiles through a combination SF induced shocks and sheer, that create non-axisymmetric motion (i.e. outflows) above and below the bar. In contrast, the \Ha velocity dispersion of NGC\,4790 is much more uniform, with lower values ($\sigma_{H\alpha}$ $\lesssim$ 40\,\kms) throughout the disc.

\begin{figure*}
    \centering
    \includegraphics[width = 0.9\textwidth]{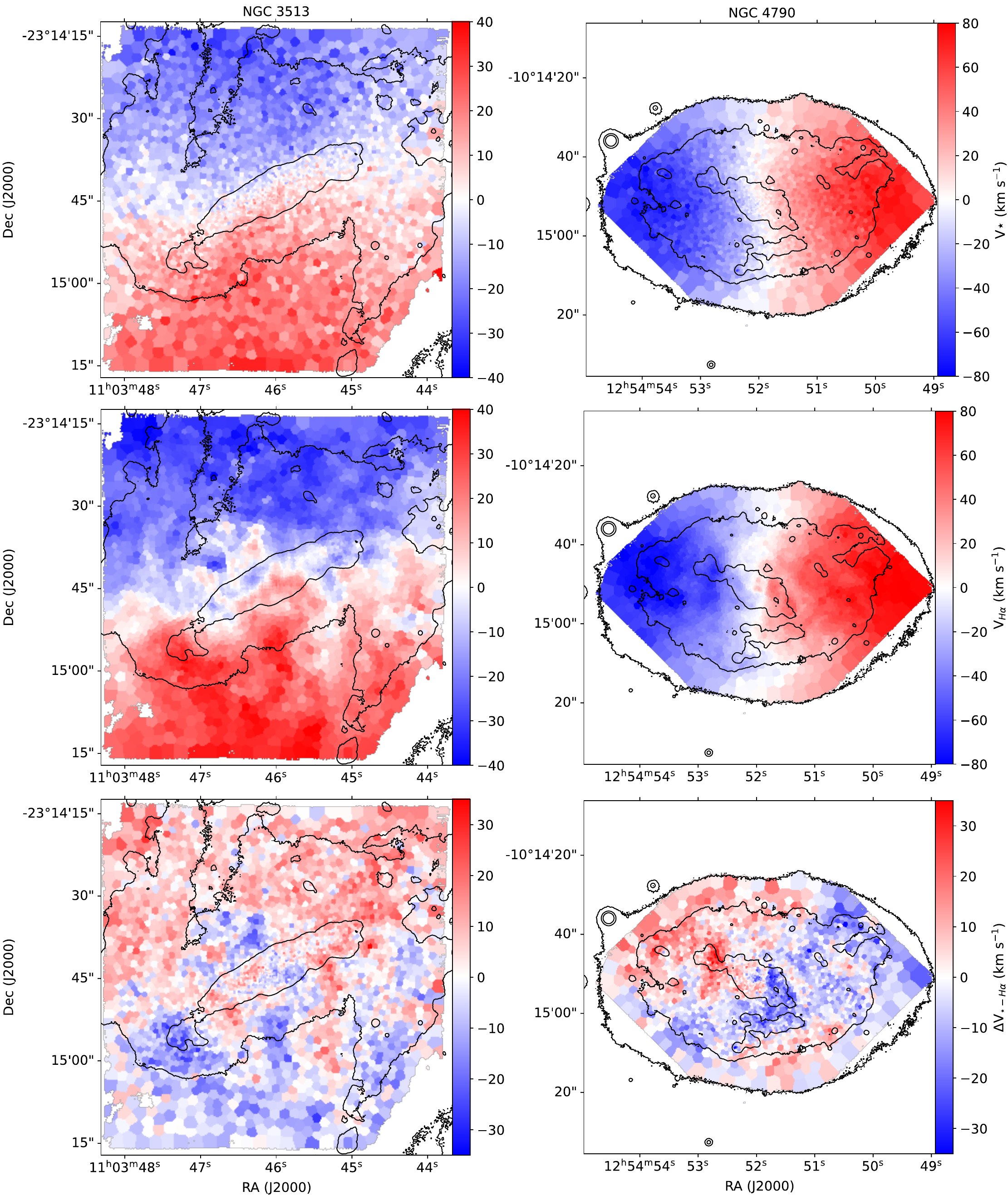}

\caption{Stellar and ionised gas velocity fields for NGC\,3513 (left) and NGC\,4790 (right). \textit{Top row}: Binned stellar velocity field. \textit{Middle}: Binned \Ha velocity field. \textit{Bottom}: Difference between them ($\Delta{V}_{\star - H\alpha}$). The black contours are the same surface brightness contours shown in Fig.\,\ref{fig:MUSE_Ha_maps}. Ordered rotation can be seen in the stars and gas of both galaxies, though non-circular orbits are clearly seen in the gas above and below the bar of NGC\,3513.}

\label{fig:MUSE_stellar_Ha}
\end{figure*}

We present the binned stellar and \Ha velocity field, and the difference between the two ($\Delta{V}_{\star - H\alpha}$) in Fig.\,\ref{fig:MUSE_stellar_Ha}. The stellar velocity field of NGC\,3513 shows regular rotation, whereas the effect of non-circular orbits above and below the bar is visible in the \Ha velocity field and co-spatial with the high \Ha velocity dispersion. The stellar and \Ha velocity field of NGC\,4790 are very similar and show a rotating disc with no significant deviations from ordered motion. The values of $\Delta{V}_{\star - H\alpha}$ difference for NGC\,4790 (right bottom panel in Fig.\,\ref{fig:MUSE_stellar_Ha}) are low and the structure is likely explained by asymmetric drift \citep{Bershady2024}. 

\section{The origin of the disturbed \HI}
\label{sec:origin}
There are some clear trends seen in the close interacting systems: i) the newly detected, diffuse \HI show the galaxies have recently been and are currently interacting; ii) the global properties (e.g. morphology \MHI, SFR, etc.) are indistinguishable from main sequence star-forming galaxies and, thus, unaffected by the interactions; and iii) the multiphase ISM (\HI, H$_{2}$ and \Ha) and stellar component of the star-forming disc are largely unaffected by the interactions, with only minor asymmetries present. 

The systems analysed here are not expected to be part of massive, X-ray emitting galaxy groups that are embedded in a dense, hot medium. Also, from their stellar mass and rotation curve, we expect that NGC\,3511 and NGC\,4781 have halo masses of a few $\times$ 10$^{11}$\,\Msol; hence, it is unlikely that their circumgalactic medium (CGM) is sufficiently dense and extended for ram pressure to be an effective environmental mechanism. Furthermore, the extended \HI morphology and velocity structure does not resemble ram pressure tails and we can likely attribute the origin of the disturbed \HI to the interaction of tidal fields. Below, we characterise the details of the previously unseen, mild-strength galaxy interactions that are only detectable below \NHI = 10$^{19}$\,\cm. 

\subsection{The history and timescale of the interactions}
It is well established that a fly-by can produce high column density \HI bridges and tails through strong gravitational interactions \citep[e.g.][]{Pearson2016, Pearson2018, Luber2022}. The gas is initially compressed during pericentric passage, and tidal features (such as bridges and tails) form along the direction of encounter after pericentric passage, when the weaker bound \HI is pulled out of the disc. It naturally follows that similar features are produced in the low column density regime, with weaker tidal perturbations. As the galaxies separate, a substantial fraction of the stripped \HI can fall back toward each galaxy and produce puffed-up \HI halos. If the galaxies undergo another passage, the change in orbital direction may cause the returning gas to lag and give rise to counter-tails \citep{Toomre1972, Hibbard1995}. 

{Using the dark matter halo mass reported in \citet{deBlok2024}, we estimated $v_{\rm circ}$ (i.e. the characteristic orbital speed of a test particle at radius, $R$, within the halo) at the location of NGC\,3513. As the $v_{\rm circ}$ of 170\,\kms is slightly below the halo escape velocity of 220\,\kms and the projected separation is less than the virial radius, it is plausible that NGC\,3513 is bound to NGC\,3511. The complex kinematics beyond the rotating discs identified in the previous section may arise from different stages of the pair interactions. As there are multiple kinematically distinct features, it is unlikely that they could be produced from a single pass. We can attribute that the main component of the bridge is most likely \HI being stripped from NGC\,3513, since it is the lower mass of the pair, making it easy to remove gas from the disc during a fly-by, and there is a smooth velocity gradient starting from the galaxy and continuing into the bridge. The extended emission to the north of NGC\,3511 is lagging with respect to the rotating disc, and the emission to the south of NGC\,3513 has a sharp change along the line-of-sight direction (see Fig.\,\ref{fig:snapshot} and the 3D renderings). These features resemble counter-tails that require a change in orbital direction and they cannot be produced at the same time as the bridge. The origin of the clouds in the arc to the east of NGC\,3511 and bridge is less clear. They may be a remnant from an additional fly-by at a epoch different to the bridge and counter-tails. There is a tentative velocity continuity when tracing the structure of the clouds, along the outskirts of the eastern emission of NGC\,3511 to the northern counter-tail, while another possibility is that they are part of the same structure. There is a lower column density \HI component to the arc of clouds (see Fig.\,\ref{fig:overlay}), although the lower spatial resolution and beam smearing does not provide additional clarity on the underlying structure.

We estimated the tidal radius (\rT), which is the point where the gravitational acceleration from an external system exceeds the galaxy's binding forces \citep{Binney1987}. When assuming point-like masses and circular orbits \citep[e.g.][]{Gadja2016} the tidal radius is \rT = [$m$/$M$/(3 + $m$/$M$)]$^{1/3}$ $\times$ $R$, with $M$ being the mass of the host galaxy, $m$ as the mass of the satellite and $R$ is the distance between the two. Using the dynamic mass (Table\,\ref{tab:global_properties}), \rT = 12\,kpc for NGC\,3511 and is shown as a dashed grey circle in Fig\,\ref{fig:mom0}. Only the high density \HI in the stellar disc is confined by the tidal radius, and \HI beyond this radius has the potential to be stripped out of the halo. The most recent pericentric passage likely occurred approximately 250\,Myr ago, where only diffuse gas (\NHI = 10$^{18.6}$\,\cm) has been pulled out of the disc beyond the tidal radius.

The NGC\,4781 field likely shows galaxies at different stages of fly-bys, although there are some similarities between the NGC\,3511/3513 pair, and NGC\,4790-UGCA\,308 interaction. The dense component of the bridge is most likely \HI being stripped from the lower mass UGCA\,308 from a recent fly-by, as there is a smooth velocity gradient originating from the galaxy that continues along the tail that points towards NGC\,4790. NGC\,4790 may have tidally perturbed NGC\,4781 from a fly-by at a different epoch. The more disturbed \HI of NGC\,4781 is present in the north (i.e. the same side of NGC\,4790) and while there is no \HI that connects them at this time, there may have been (e.g. a bridge) in the past. This indicates that the NGC\,4781-4790 fly-by may have occurred first and the NGC\,4790-UGCA\,308 fly-by occurred soon after. If this is the case, then NGC\,4790 is closer to a pericentric passage fly-by, where gas compression is at the highest, which could explain the extended \HI emission along the major axis. By comparison, NGC\,4781 would then be in a later stage past pericentric passage and, hence, there has been more time for the gas pulled out of the disc to re-settle and produce a puffed up halo of \HI. 

Using the NGC\,4781 dark matter halo mass \citep{deBlok2024}, the $v_{\rm circ}$ for NGC\,4790 and UGCA\,308 are both equal to 110\,\kms. These speeds imply that pericentric passage for the NGC\,4790-UGCA\,308 fly-by was approximately 400\,Myr ago, the NGC\,4781-NGC\,4790 fly-by occurred and 500\,Myr ago, consistent with the proposed timeline. Using the dynamic masses (Table\,\ref{tab:global_properties}), the NGC\,4781 \rT = 30\,kpc and was computed with NGC\,4790 as the satellite, and the NGC\,4790 \rT = 16\,kpc, and was computed with UGCA\,308 as the satellite. Both tidal radii are shown as a dashed grey circle in Fig\,\ref{fig:mom0}, where approximately all of the disturbed \HI of NGC\,4781 (other than a few clouds) is contained within its \rT, and nearly all of the disturbed \HI of NGC\,4790 is contained within its \rT. We are not able to identify a clear origin of the clouds in the IGM, although it is unsurprising to see some debris not associated with a galaxy, given the recent tidal activity.

The complexity of the \HI distribution and kinematics make it unlikely that a single fly-by could reproduce what is observed in either system. Tidal forces appear to have pulled only the diffuse \HI out of the disc, which is clearly not in equilibrium. The strength of the tidal interaction is weak as the dense gas remains close to the disc, and the mass transfer must be low. Given the low orbital speeds within the halo, these galaxies have likely been interacting with mild tidal forces for the past $\sim$ gigayear. As the rotation velocity of the \HI discs are on the same order of magnitude as the velocities within the halo, there is enough time for the majority of the \HI to settle back onto the disc \citep{Holwerda2011}. This process may correlate with the amount of \HI beyond the tidal radius, where the galaxies (NGC\,3511 and NGC\,4790) with the more recent fly-by show more \HI beyond the tidal radius, compared to the galaxy (NGC\,4781) that has had more time for the disturbed gas to settle. The potential keeps the diffuse \HI bound, combined with the angular momentum of the gas, creates coherent gas flows that are wrapping around the stellar discs (Fig.\,\ref{fig:vel_field}), an intermediate stage of the gas re-settling. These arcs are clear in NGC\,3511, NGC\,4781 and NGC\,4790, the more inclined late types with higher rotation velocities. There is a hint of this coherent gas arc in NGC\,3513, although it is tentative as a result of the galaxy being face-on with a low rotation velocity. This combined effect may also explain why there is an offset in the kinematic major axis of the \HI at different densities.

To search for kinematically anomalous \HI within the disc region and quantify the non-rotating components at large radii, we create a \BBarolo locally normalised model of each galaxy. This procedure normalises the flux density in each pixel, and is a more accurate representation of the \HI structure in the data compared to using a constant density in each ring from the azimuthally averaged model. We subtract the locally normalised model from the data to identify \HI emission at forbidden velocities and \HI beyond the outer ring used in the \BBarolo fitting, determined by the radius where there is enough signal for the model to converge to sensible values. The remaining emission ranges between 4 and 18\%, where the largest (11 and 18\%) are associated with UGCA\,308 and NGC\,4790 respectively, and is consistent with these galaxies having the most recent fly-bys, and the least amount of time for the \HI to settle back onto the disc.

\HI tidal features are commonly observed in interacting galaxies \citep[e.g.][]{Koribalski2004a, deBlok2018, Namumba2021, Wang2023, Staveley-Smith2025}. Due to sensitivity limitations, most of the previous cases in the literature observe \HI tidal features above \NHI = 10$^{19.5}$\,\cm. In this work, the prominent \HI tidal features have typical column density in the 10$^{18.5}$ -- 10$^{19}$\,\cm range. The \HI mass of the tidal features are between log(\MHI/\Msol) = 7.0 and 7.6, and only contribute approximately 2\% to the total \MHI of the systems. Up until deep, resolved observations from the MeerKAT and FAST telescopes, these low-column density tidal features were below the sensitivity limit. \citet{Wang2024} find that more diffuse \HI is found in galaxies with prominent tidal-interaction signatures, and is consistent with our findings -- that there is always a diffuse \HI component to the higher column density tidal features. 

The close systems in this work are analogous to low-mass versions of the M81 triplet \citep{deBlok2018}. The \HI tidal features (bridges and tails) in the close systems are a factor of 10 -- 100 more diffuse (and contribute less mass) than what is seen in the M81 triplet. Indeed, the shallower potentials result in slower fly-bys that leave the star-forming disc and high column density \HI in the ISM largely undisturbed (e.g. Figs.\,\ref{fig:Rmol_maps} and \ref{fig:MUSE_Ha_maps}). Even though slower speeds result in longer tidal encounters, the stellar and gaseous components have enough time to respond and adapt to the changing potential. This results in the outer \HI envelope with longer orbital period and lower binding energy to be perturbed and stripped if it lies beyond the tidal radius, while the internal structure of the galaxy with shorter orbital periods is less perturbed \citep{Binney1987}. The kinematics of the different components of the star-forming disc all display ordered rotation and there has not been substantial gas compression or shocks to induce a star-burst. There is no stellar component to any \HI tidal feature down to a surface brightness of $\mu_{g}$ = 29\,mag\,arcsec$^{-2}$. Using the Subaru telescope, a star count analyses of the M81 triplet revealed a very low surface brightness stellar component associated with the \HI tidal features \citep{Okamoto2015}. We cannot rule out the possibility of very faint tidally stripped stars beyond our surface brightness sensitivity in the close systems, although optical images a few mag\,arcsec$^{-2}$ deeper than the VST images are needed to determine this. If there is a very low surface brightness stellar component to the \HI tidal features, it is unlikely to contribute significantly to the stellar mass in these systems as the interactions are weaker than what is seen in the M81 triplet.

\subsection{The dwarf irregulars 2MASX\,J1255 and MKT\,25413}
While we have largely focussed on the interacting triplet in the NGC\,4781 field, it is worth discussing the properties of the two dwarf irregular galaxies detected with \HI: 2MASX\,J1255 and MKT\,25413. There is no \HI detected between NGC\,4781 and 2MASX\,J1255, and the projected distance between them is sufficiently large (114\,kpc) that clearly these galaxies are not interacting. The asymmetric and lopsided stellar body observed in 2MASX\,J1255 (Fig.\,\ref{fig:overlay}) may be due to the combination of stochastic SF and a shallow potential well, which are common in low-mass galaxies \citep[e.g.][]{vanZee2000, vanZee2001,Lazar2024, Lazar2026}. There has not been substantial gas loss or quenching for 2MASX\,J1255, as it lies on the \FHI gas fraction and sSFR main sequence (Fig.\,\ref{fig:scaling}).

As previously mentioned, at a distance of 11.3\,Mpc and log(\MHI/\Msol) = 5.7, MKT\,25413 has the lowest \MHI detected of a galaxy that is not a Milky Way satellite. Other galaxies with comparable \HI masses are the low surface brightness dwarf (LSB-D) in the Dorado group with log(\Mstar/\Msol) and log(\MHI/\Msol) = 6.3 and 6.1 at a distance of 17.7\,Mpc \citep{Maccagni2024a}, the isolated galaxy Pavo with log(\Mstar/\Msol) and log(\MHI/\Msol) = 6.1 and 5.8 at a distance of 2.2\,Mpc \citep{Jones2025}, and the Milky Way Satellite Leo\,T with log(\Mstar/\Msol) and log(\MHI/\Msol) = 5.3 and 5.6 \citep{Adams2018}. Even with only a few known galaxies of this mass with \HI, they are in different environments and display a range of gas fractions. For example, LSB-D is a gas-rich low surface brightness galaxy, with some recent SF and is possibly rotation supported. In comparison, MKT\,25413 has a similar stellar density and environment; however, it is extremely gas-poor, with no detectable SF (Fig.\,\ref{fig:scaling}, Table\,\ref{tab:global_properties}). MKT\,25413 has the same \FHI as Leo\,T, although it is likely that Leo\,T has a higher \MHI. Due to its large angular extent it is possible not all flux has been detected \citep[the zero spacing problem; see e.g.][]{Wang2024}. \citet{Jones2025} has shown that, a galaxy of this mass with a \HI reservoir can be at different stages of the boom-and-bust SF cycle. As MKT\,25413 is such a low-mass galaxy and is in close proximity to NGC\,4781, it is likely to have been influenced via ram-pressure where some CGM is expected to be present, or from the tidal fields. The lack of recent SF may imply a low SF efficiency, placing the galaxy in the `bust' part of the cycle. Another possibility is the \HI has been captured from an external source, such as the adjacent diffuse \HI from NGC\,4781 -- a similar scenario to that of FCC\,248 and the \HI tail of NGC\,1427A \citep{Kleiner2023}. 

\subsection{Comparison to simulations}
Both N-body and hydrodynamical simulations are able to reproduce the galaxy \HI-environmental dependence -- whereby tidal interactions and ram-pressure stripping remove gas and suppress SF \citep[e.g.][]{Tonnesen2009, Marasco2016, Steinhauser2016, Chen2024}. Simulations additionally show the diffuse, low-density gaseous component extends the farthest, and is therefore more sensitive to environmental stripping and tidal disturbances \citep[e.g.][]{Tonnesen2009, Steinhauser2016, vandeVoort2019, Marasco2025}. However, most theoretical work has focussed on merging galaxies and gas stripping in dense, cluster environments. 

The recent \textsc{NewHorizon} and \textsc{NewHorizon2} cosmological hydrodynamic zoom-in simulations \citep{Dubois2021, Yi2024} are able to explore the stellar and gas properties of non-merging, low-mass (i.e. log(\Mstar/\Msol) $<$ 10) galaxies in low-mass group halos. These simulations resolve spatial scales down to 34 and 68\,pc and masses of $M$ $\sim$ 1 -- 2 $\times$ 10$^{4}$\,\Msol, which can measure the global properties and resolve the stellar and gas components of galaxies with log(\Mstar/\Msol) $>$ 7. This is a very good match with the galaxies and environment in this work and we compare our results to the quenching and gas-stripping processes for satellite galaxies in low mass groups in \citet{Rhee2024} and \citet{Rhee2026}. A small caveat is the simulated halo (and thus central galaxy) masses are slightly more massive than ours. Furthermore, \citet{Rhee2026} does not probe below \NHI = 10$^{19}$\,\cm. 

The quenching of satellite galaxies was found to be a complex interplay between a variety of mild environmental effects \citep{Rhee2024}. The combination of SF efficiency and availability of dense gas are the two most important factors, and both depend on the orbital motion and mass of the satellite. The satellite galaxies with log(\Mstar/\Msol) $>$ 9.1 continue to form stars for several orbital times, and galaxies with log(\Mstar/\Msol) $>$ 8.2 retain their ISM gas for extended periods \citep{Rhee2024}. This matches our proposed scenarios for the interactions presented in this paper. \citet{Rhee2024} pointed out that the sustained SF can include multiple rejuvenation phases following pericentric passage, which is only possible when the cold gas reservoir is disturbed but not stripped, similar to what we see in this work. An example may be UGCA\,308 slightly below the star-forming main sequence, due to a lower SF efficiency right after the fly-by with NGC\,4790. 

The findings of \citet{Rhee2026} are remarkably consistent with the mild level of tidal interactions inferred in this work. In their simulations, the majority of satellite galaxies retain substantial gas reservoirs over long timescales, while showing signs of outer gas stripping. This behaviour is more prevalent among higher-mass satellites, a regime in which NGC\,3513 and NGC\,4790 reside. Although these satellites remain gas rich, most exhibit noticeable gas-tail morphologies, indicating that environmental effects within group halos are insufficiently strong to remove the main body of the ISM \citep{Rhee2026}. Furthermore, even at peak interaction strength (i.e. pericentric passage), the external pressure is unable to strip the ISM, but is capable of disrupting and reshaping the diffuse gas, while largely leaving the stellar distributions and total gas masses unaffected.

While the main findings of \citet{Rhee2026} are highly consistent with our results, the least massive central galaxy in their group sample has log(\Mstar/\Msol) = 10.2 (their Table\,1), which is more massive than NGC\,3511 and NGC\,4781. Their environmental effects may therefore be somewhat stronger than those probed in our systems. This is reflected in the more prominent simulated gas tails seen at \NHI $>$ 10$^{19}$\,\cm, which are not as pronounced in our data. We note, however, that their gas tails are shown near pericentric passage, whereas our systems are likely observed after pericentric passage, which may have allowed some of the disturbed dense gas to re-settle. Finally, \citet{Rhee2026} find that satellites with log(\Mstar/\Msol) $<$ 8 are more easily stripped of their gas (via multiple pathways). As 2MASX\,J1255 is not gas poor, it is unlikely to have interacted with the other group galaxies thus far.

High $\sigma_\mathrm{HI}$ at large radii with complex \HI line profiles have been reported in the TNG50 and FIRE-2 simulations by \citet{Marasco2025}. These features were partly attributed to pristine gas accretion and even though this differs to the tidal disruptions we observe in our galaxies, the phenomenon may be the same. In both cases, there is an external gas flow circulating and mixing with the periphery of the ISM, where the gas is in the process of falling onto the disc. The peak $\sigma_\mathrm{HI}$ of the simulated galaxies reaches $\sim$ 60\,\kms, a similar value to what we observe. A difference is that the high $\sigma_\mathrm{HI}$ regions in the outer discs of the TNG50 and FIRE-2 simulations are not associated with SF, while the regions with high $\sigma_\mathrm{HI}$ associated with the galactic fountain process are \citep{Marasco2025}.

\section{Summary}
\label{sec:Summary}
We present a multi-wavelength analysis of the close interacting systems in the MHONGOOSE survey. Using the deepest and most resolved \HI images of nearby star-forming galaxies to date, we detected \HI in seven galaxies: two in the NGC\,3511 field and five in the NGC\,4781 field. The \HI below \NHI = 10$^{19}$\,\cm reveals tails and bridges that connect NGC\,3511 and NGC\,3513 as an interacting pair, as well as puffed-up \HI, tails, and bridges related to interactions between NGC\,4781, NGC\,4790, and UGCA\,308. The two additional galaxies detected with \HI in the NGC\,4781 field are two dwarf irregulars; 2MASX\,J1255 and MKT\,25413, where the latter has the lowest \MHI detected in a galaxy that is not a Milky Way satellite. 

Six of the seven galaxies lie on (or within) the scatter of the \FHI gas-fraction and star-forming main sequences. The exception is MKT\,25413, which is substantially below both scaling relations and most noticeably in the sSFR relation, with no detectable signs of recent SF. The morphology of the extended \HI features of the interacting galaxies is best explained by tidal fields drawing the diffuse \HI out of the disc from multiple fly-bys in the NGC\,3511/3513 pair, and with NGC\,4790 first passing NGC\,4781 and UGCA\,308 flying by NGC\,4790 shortly after. We find a tight correlation between the tidal radius of the galaxy and the extent of the disturbed low column density \HI, implying that the bulk of the \HI remains gravitationally bound to each respective galaxy and only a small amount may have been removed at $r$ $>$ \rT.

Through quantifying the stellar and multiphase gaseous components of the ISM, we find that the four most massive, late type galaxies have ordered rotation within the stellar disc, with only minor asymmetries present. The kinematic modelling of NGC\,3511 and NGC\,4781 show that \HI and H$_{2}$ are co-rotating, and small differences are due to the fact that H$_{2}$ is dynamically colder. Both galaxies contain typical atomic-to-molecular gas ratios of star-forming galaxies. The MUSE observations detect \Ha flux in the discs of NGC\,3513 and NGC\,4790, and is prominent in the bar of NGC\,3513, indicating ongoing SF in these structures. The stellar and ionised gas velocity asymmetries in the central region of NGC\,3513 indicate ongoing gas flow along the bar.

In this work, deep \HI observations have revealed mild tidal interactions that have previously remained unseen. There is no stellar component in the VST deep optical images associated with any of \HI tails, bridges, or clouds in the IGM, which confirms the gentle nature of the interactions. The consequence of these interactions is to redistribute the diffuse \HI without stripping it, while leaving the kinematics of the stellar disc and ISM unaffected. Therefore, the galaxies retain an abundant cold gas reservoir and the baryon cycle is mostly undisturbed. This work is limited to just a handful of galaxies, although we expect many systems like this to be revealed with the next generation of mid-frequency radio telescopes, such as DSA, ngVLA, and SKA-mid. When they come online, a much larger sample of close galaxy systems will be observed below \NHI = 10$^{19}$\,\cm, providing a statistical picture of how mild environmental interactions impact the evolution of galaxies.

\section{Data availability}
\label{sec:avail}
The observations presented in the paper are part of the MHONGOOSE survey \citep{deBlok2024}. These data are publicly available via the MHONGOOSE website (\url{mhongoose.astron.nl}). 

\begin{acknowledgements}
The MeerKAT telescope is operated by the South African Radio Astronomy Observatory, which is a facility of the National Research Foundation, an agency of the Department of Science and Innovation. We are grateful to the full MeerKAT team at SARAO for their work on building and commissioning MeerKAT. This work used data collected with the INAF VST telescope at the ESO Paranal Observatory. This work made use of the Inter-University Institute for Data Intensive Astronomy (IDIA) visualisation lab (https://vislab.idia.ac.za). IDIA is a partnership of the University of Cape Town, the University of Pretoria and the University of Western Cape. This work is partly based on observations collected at the European Organisation for Astronomical Research in the Southern Hemisphere under ESO programmes 097.B-0165 and 098.B-0551. This work has received funding from the European Research Council (ERC) under the European Union’s Horizon 2020 research and innovation programme (grant agreement No. 882793 "MeerGas"). JH acknowledges support from the UK SKA Regional Centre (UKSRC). The UKSRC is a collaboration between the University of Cambridge, University of Edinburgh, Durham University, University of Hertfordshire, University of Manchester, University College London, and the UKRI Science and Technology Facilities Council (STFC) Scientific Computing at RAL. The UKSRC is supported by funding from the UKRI STFC. FMM carried out part of the research activities described in this paper with contribution of the Next Generation EU funds within the National Recovery and Resilience Plan (PNRR), Mission 4 - Education and Research, Component 2 - From Research to Business (M4C2), Investment Line 3.1 - Strengthening and creation of Research Infrastructures, Project IR0000034 – ``STILES - Strengthening the Italian Leadership in ELT and SKA''. EDT was supported by the European Research Council (ERC) under grant agreement no.\,10104075.

\end{acknowledgements}

\bibliographystyle{aa} 
\bibliography{References}

\begin{appendix} 
\section{The optical and \HI emission of MKT\,25413-102444}
\label{sec:appen_MKT}
The evidence supporting the association of the stellar and \HI components of MKT\,25413 are as follows; the spatial coincidence of the stellar and \HI emission, and the difference in \HI velocity with respect to the adjacent emission. In Fig\,\ref{fig:MKT_overlay} we show a zoom-in of Fig\,\ref{fig:overlay}, which overlays the \NHI = 10$^{18.7}$ and 10$^{17.9}$\,\cm emission (red and black contours from the \texttt{r10\_t00} and \texttt{r05\_t60} cubes, respectively) on the VST $g$-band optical image. The stellar emission is approximately centred within the beam of the higher resolution \HI emission, which is the exact location expected for a galaxy that hosts a \HI reservoir. At lower column densities, the moment maps of the full NGC\,4781 field (see the figures in the main text) appear to show that MKT\,25413 is embedded in the halo of NGC\,4781. However, this is a projection effect from beam smearing and collapsing the emission along the line-of-sight. As shown in Fig.\,\ref{fig:MKT_overlay}, MKT\,25413 is detected as a unique object in the source finding procedure, not connected to the NGC\,4781 halo. While this depends on the \SoFiA settings (i.e. linking-length), the evidence is strengthened when inspecting the velocity field. 

\begin{figure}
\centering
    \includegraphics[width = \columnwidth]{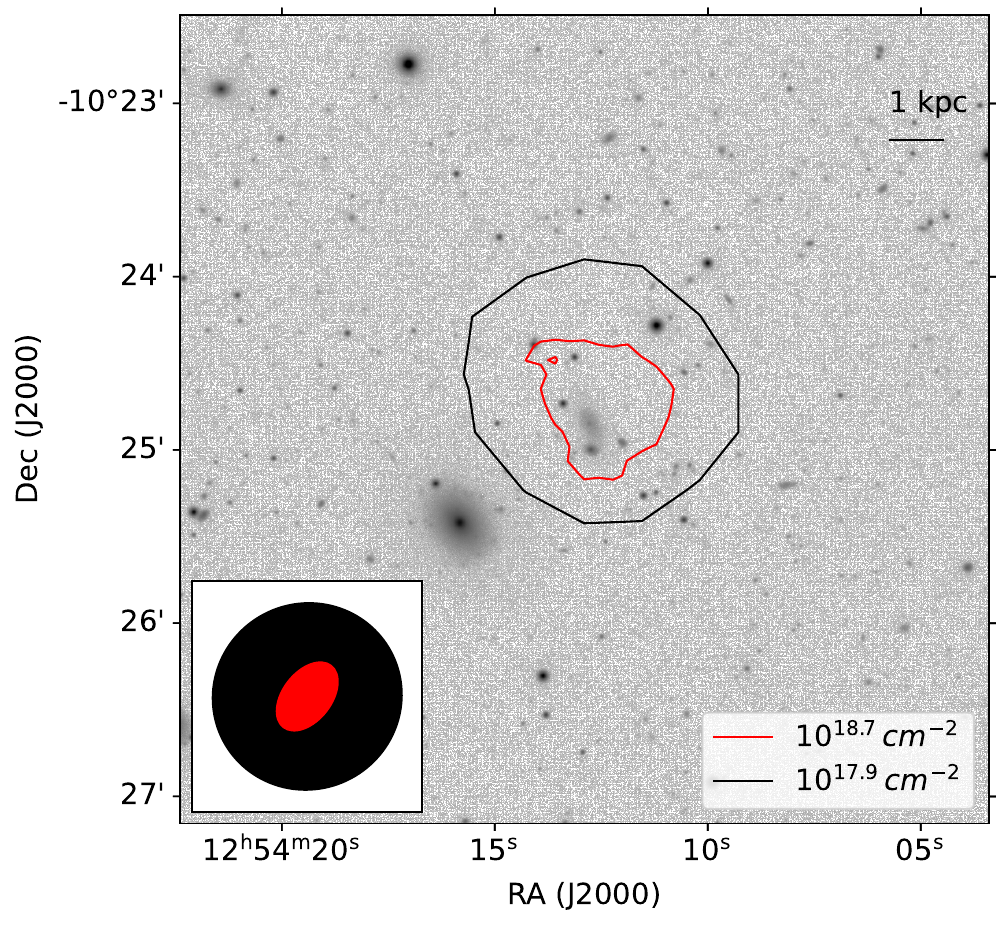}

\caption{Optical and \HI overlay for MKT\,25413: a zoom in of Fig.\,\ref{fig:overlay} that shows the VST $g$-band image overlaid with the primary beam-corrected \texttt{r10\_t00} (red) and \texttt{r05\_t60} (black) \HI contours for MKT\,25413. The 3$\sigma$ column densities over 16\,\kms are shown in the bottom-right, the synthesised beams are shown in the bottom-left with using the same contour colours, and a 1\,kpc scale bar is placed in the top-right.}

\label{fig:MKT_overlay}
\end{figure}

In Fig\,\ref{fig:MKT_vel_field}, we show the \texttt{r10\_t00} and \texttt{r05\_t60} velocity fields, centred on NGC\,4781 and re-scaled to the NGC\,4781 \vsys $\pm$ 120\,\kms. The \vsys of MKT\,25413 is 1334\,\kms, which separates it from the nearby halo gas in the \texttt{r10\_t00} image (Fig\,\ref{fig:MKT_vel_field}, top panel), and can also be seen in the channel maps between 1319 -- 1358\,\kms in the next section. In the \texttt{r05\_t60} velocity field at lower column densities, the \HI of NGC\,4781 overlaps in projection with MKT\,25413, but is separated in velocity space by $\sim$ 60\,\kms, maintaining that is it is not embedded in the halo gas (Fig\,\ref{fig:MKT_vel_field}, bottom panel). 

\begin{figure}
\centering
    \includegraphics[width = \columnwidth]{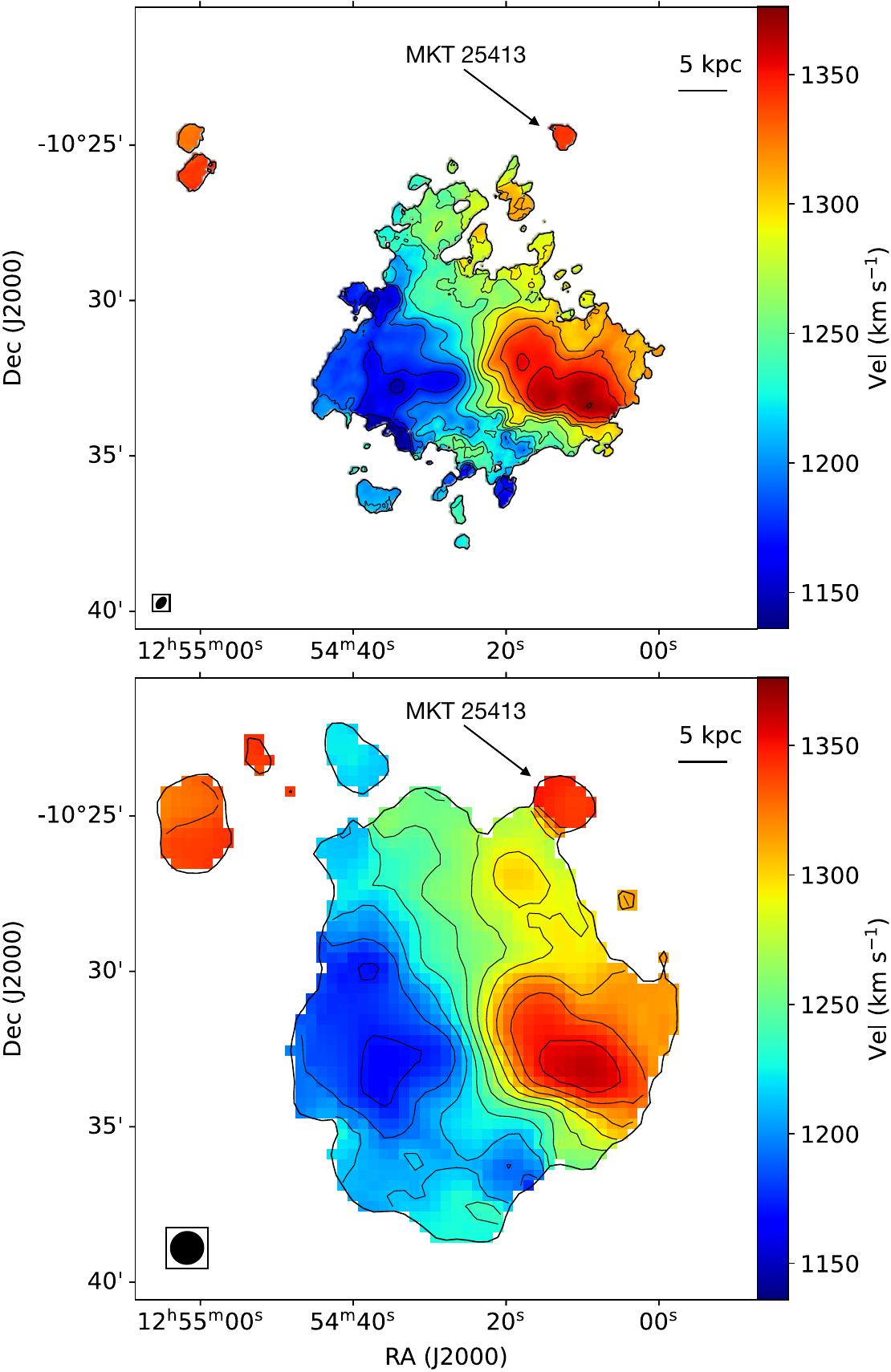}

\caption{\HI velocity field of MKT\,25413. The \texttt{r10\_t00} (\textit{top}) and \texttt{r05\_t60} (\textit{bottom}) \HI velocity field containing NGC\,4781 and MKT\,25413. The velocity scale is centred on the NGC\,4781 \vsys $\pm$ 120\,\kms, the synthesised beam of each image is shown in the bottom-left, and a 5\,kpc scale bar is shown in the top-right.}

\label{fig:MKT_vel_field}
\end{figure}

It is possible that the overlapping \HI and stellar emission is a coincidence and not physically associated with each other. The detection is in an area of \HI debris, and MKT\,25413 has the same velocity as the cloud(s) to the east in Fig.\,\ref{fig:MKT_vel_field}. The \MHI of the cloud(s) is 10$^{6.3}$\,\Msol and show a hint of a velocity gradient, whereas the \MHI of MKT\,25413 is 10$^{5.7}$\,\Msol, does not show a velocity gradient and is thus dispersion dominated. The similarity in properties to a \HI cloud with no optical emission might imply that MKT\,25413 is another cloud that is part of the debris. 

However, MKT\,25413 is similar to other low-mass, low surface brightness galaxies that have been detected with \HI in the MHONGOOSE survey. A total of four have been detecting with a \MHI below 10$^{6}$\,\Msol, with mean surface brightnesses ranging between $\mu_{g}$ = 22.5 and 24.3\,mag\,arcsec$^{-2}$. Furthermore, the MKT\,25413 $g - r$ = 0.64, which is the same as other low-surface brightness galaxies detected with \HI \citep{Honey2018}. This would lead us to favour a scenario where MKT\,25413 has a stellar and \HI component.  

\section{Resolved \HI emission in 2D and 3D}
\label{sec:appen_dist}

In this section we present a snapshot of the 3D rendering, and channel maps. A snapshot of the NGC\,3511/3513 \texttt{r10\_t00} cube, created with the \texttt{ViSL3D} software\footnote{\url{https://github.com/ixakalabadie/ViSL3D}} \citep{Labadie-Garcia2025} is shown in Fig.\,\ref{fig:snapshot}. There is a sharp velocity change along the line-of-sight in the NGC\,3511 counter tail (Fig.\,\ref{fig:snapshot}), that is easily missed in the channel maps and velocity field. The interactive renderings of all six MHONGOOSE resolutions for both systems can be found at as online material, where the units of the different surfaces are in $\sigma$. We highly encourager the reader to explore the cubes in 3D to inspect and admire the complex \HI emission. 

\begin{figure}
\centering
    \includegraphics[width = \columnwidth]{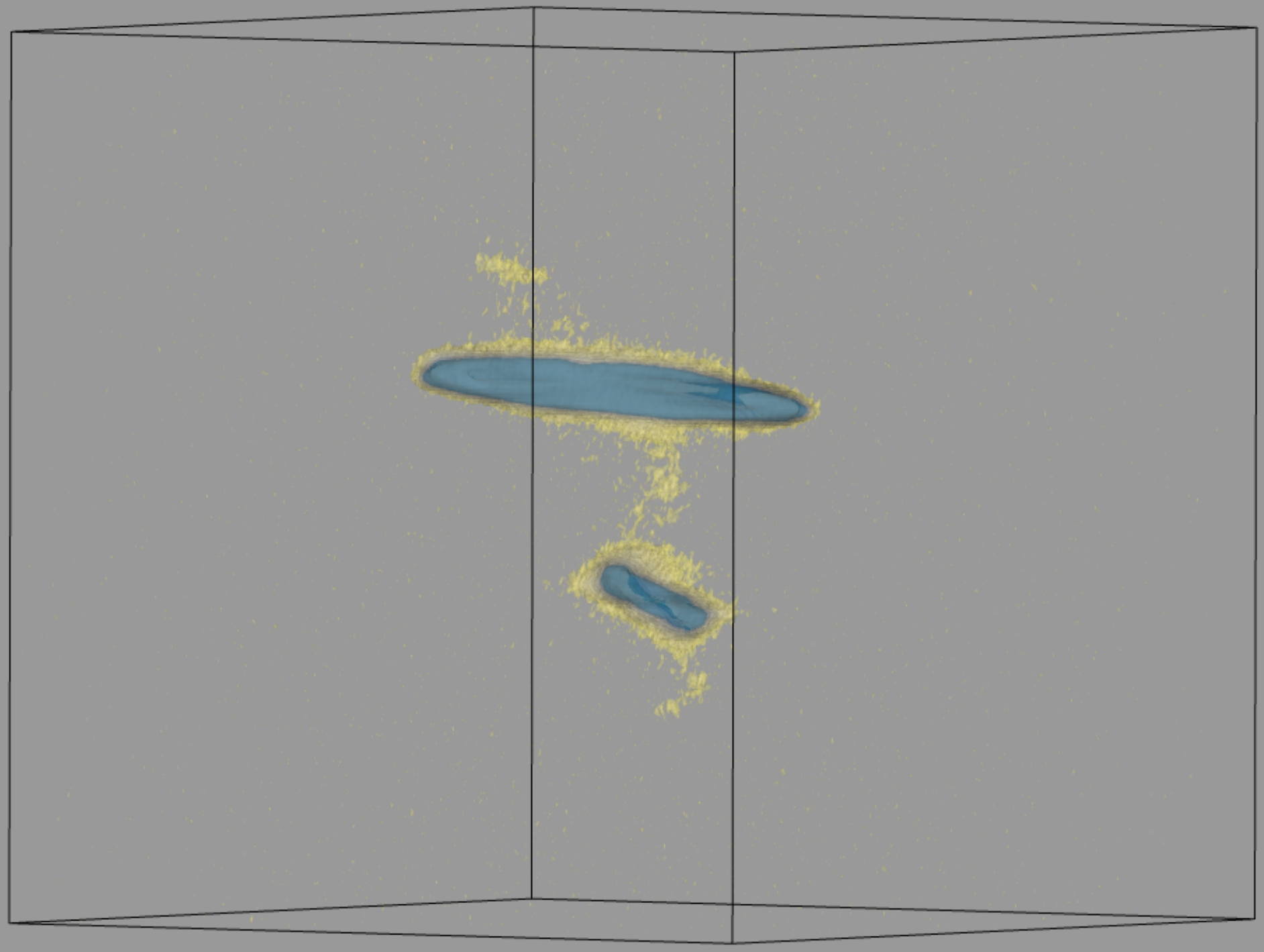}

\caption{Example rendering: snapshot of the NGC\,3511/3513 \texttt{r10\_t00} rendered cubes using \texttt{ViSL3D} \citep{Labadie-Garcia2025}. The associated interactive rendering is available online.}

\label{fig:snapshot}
\end{figure}

The \texttt{r10\_t00} channel maps, of the NGC\,3511/3513 pair and the interacting galaxies of the NGC\,4781 field binned to 10\,\kms are in Figs.\,\ref{fig:chan_map} and \ref{fig:chan_map2}, respectively. The channel maps show that a large fraction of the extended \HI beyond the stellar disc is not an extension of the inner rotating discs. The maps also show similarities between the NGC\,3513/3511 bridge and the \HI bridge connecting NGC\,4790 and UGCA\,308.

\begin{figure*}
\centering
    \includegraphics[width = \textwidth]{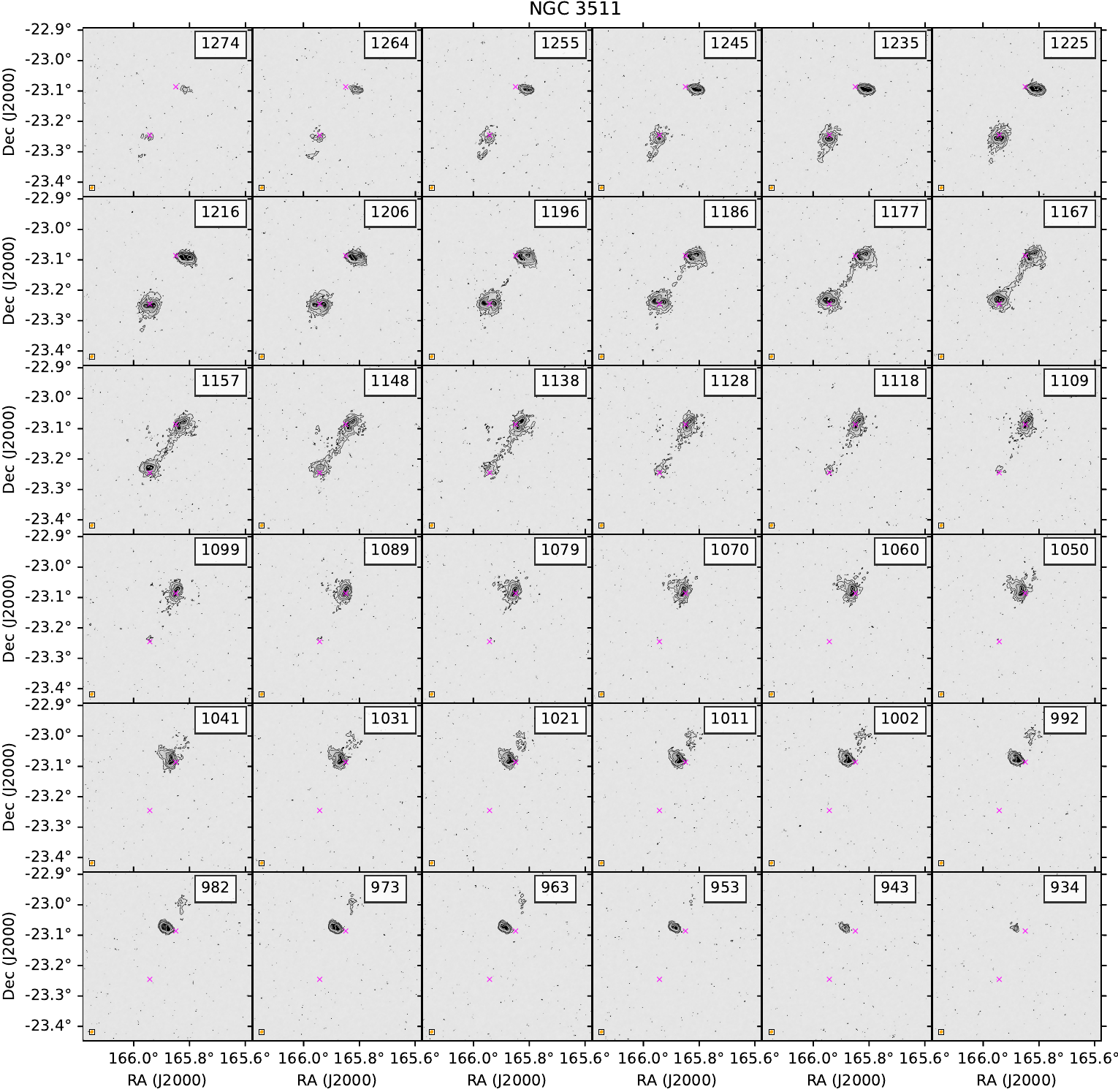}

\caption{\HI channel maps if the NGC\,3511/3513 pair. For display purposes, the channels are binned to 10\,\kms. The 3$\sigma$ contour levels start at 0.17\,mJy\,beam$^{-1}$ and increase by 4$^{n}$ with $n$ = 1, 2, 3... The optical centres of the galaxies are shown with a magenta cross. The mean velocity in units of \kms shown in the top-right corner of each channel, and the $\sim$ 28\arcsec $\times$ 17\arcsec\xspace beam is located in the bottom-left of each panel. The rotation of the late types discs are clearly seen in the channel maps, along with many additional \HI components not connected to the rotating discs.}

\label{fig:chan_map}
\end{figure*}

\begin{figure*}
\centering
    \includegraphics[width = \textwidth]{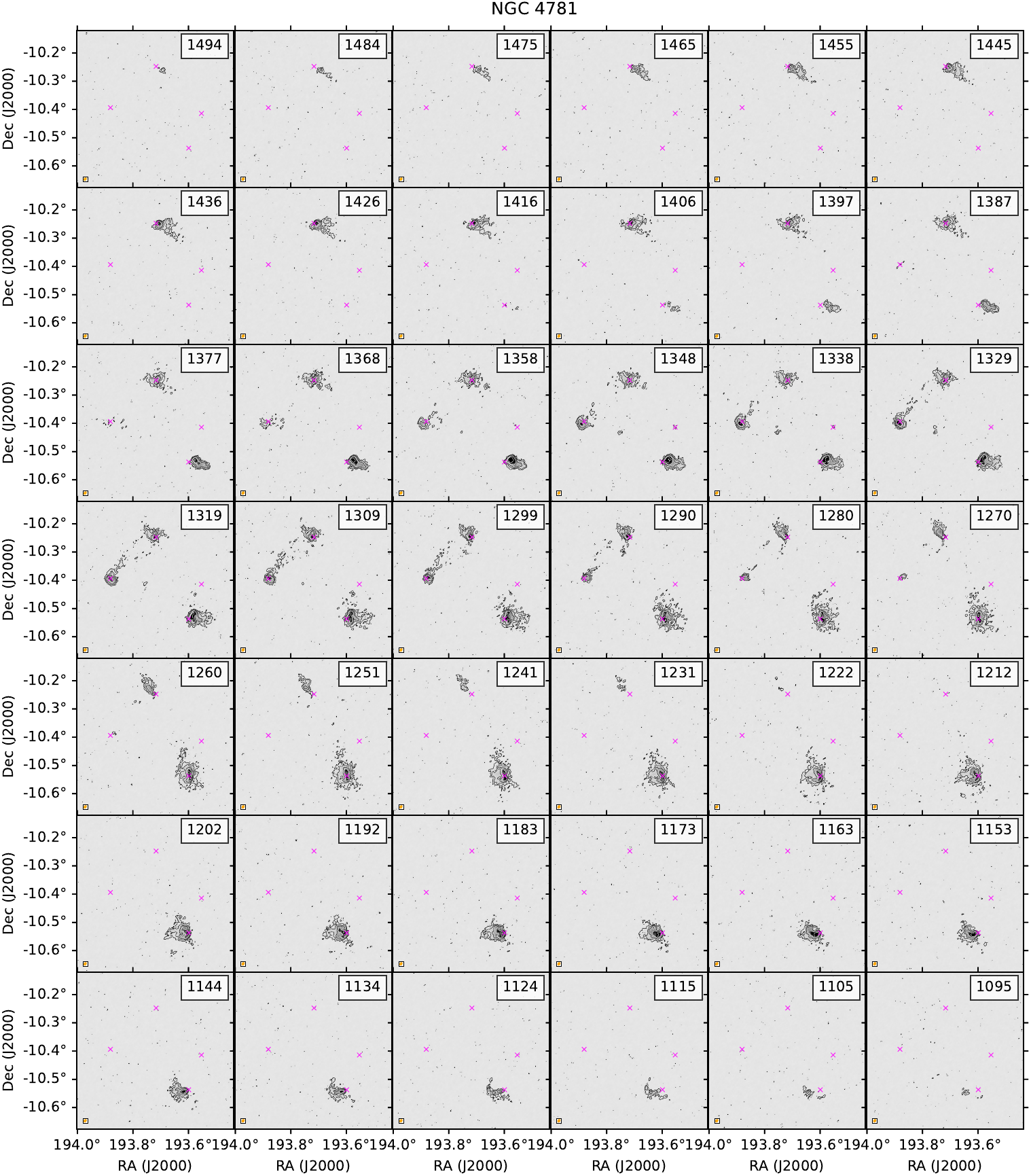}
    
   \caption{Same as Fig.\,\ref{fig:chan_map} but for the interacting galaxies in the NGC\,4781 field.}

\label{fig:chan_map2}
\end{figure*}

\end{appendix}
\end{document}